\documentclass[10pt,a4paper]{article}
\usepackage[latin1]{inputenc}
\usepackage{amsmath}
\usepackage{amsfonts}
\usepackage{amssymb}
\usepackage{graphicx}
\usepackage{xcolor}
\usepackage{lineno}

\usepackage{hyperref}
\newcommand{\footremember}[2]{%
   \footnote{#2}
    \newcounter{#1}
    \setcounter{#1}{\value{footnote}}%
}
\newcommand{\footrecall}[1]{%
    \footnotemark[\value{#1}]%
} 

\usepackage{pgfplots}
\pgfplotsset{compat=newest}
\usetikzlibrary{plotmarks}
\usetikzlibrary{arrows.meta}
\usepgfplotslibrary{patchplots}
\usepackage{grffile}

\newlength\figureheight
\newlength\figurewidth

\usepackage[left=1.5cm,right=1.5cm,top=1.5cm,bottom=1.5cm]{geometry}

\author{Cynthia Michalkowski\footremember{note2}{Robert Bosch GmbH, Center for Research and Development, Robert-Bosch-Campus 1, 71272 Renningen, Germany }, Maziar Veyskarami\footremember{note1}{Institute for Modelling Hydraulic and Environmental Systems, University of Stuttgart, Pfaffenwaldring 61, 70569 Stuttgart, Germany}, Carina Bringedal\footrecall{note1}, Rainer Helmig\footrecall{note1}, Veronika Schleper\footrecall{note2}}
\title{Two-phase flow dynamics at the interface between GDL and gas distributor channel using a pore-network model}

\date{January 2022}
\begin{document}
	\maketitle
	\section*{Abstract}
	For improved operating conditions of a polymer electrolyte membrane (PEM) fuel cell, a sophisticated water management is crucial. Therefore, it is necessary to understand the transport mechanisms of water throughout the cell constituents especially on the cathode side, where the excess water has to be removed. Pore-scale modeling of diffusion layers and gas distributor has been established as a favorable technique to investigate the ongoing processes.\\
    Investigating the interface between the cathode layers, a particular challenge is the combination and interaction of the multi-phase flow in the porous material of the gas diffusion layer (GDL) with the free flow in the gas distributor channels. The formation, growth and detachment of water droplets on the hydrophobic, porous surface of the GDL have a major influence on the mass, momentum and energy exchange between the layers.\\
    A dynamic pore-network model is used to describe the flow through the porous GDL on the pore-scale. To capture the droplet occurrence and its influence on the flow, this dynamic two-phase pore-network model is extended to capture droplet formation and growth at the surface of the GDL as well as droplet detachment due to the gas flow in the gas distributor channels. In this article, the developed model is applied to single- and multi-tube systems to investigate the general drop behavior. These rather simple test-cases are compared to experimental and numerical data available in the literature. Finally, the model is applied to a GDL unit cell to analyse the interaction between two-phase flow through the GDL and drop formation at the interface between  GDL and gas distributor channel.
    \section*{Article Highlights}
pore-network model, two phase flow, droplet formation, droplet interaction
\section*{Declarations}
\subsection*{Funding}
This work was funded by the Robert Bosch GmbH and supported by the Deutsche Forschungsgemeinschaft (DFG, German Research Foundation) by funding SFB 1313, Project Number 327154368.
\subsection*{Conflicts of interest}
The authors declare that they have no conflict of interest.
\subsection*{Code availability}
All code is available online in the Dumux gitlab repository (https://git.iws.uni-stuttgart.de/dumux-pub/michalkowski2022a)
\subsection*{Availability of data and material}
All relevant data can be generated with the open source code.
\subsection*{Authors' contributions}
All authors contributed to the study conception and design. Material preparation, data collection and analysis were performed by Cynthia Michalkowski and Maziar Veyskarami. The first draft of the manuscript was written by Cynthia Michalkowski and all authors commented on previous versions of the manuscript. All authors read and approved the final manuscript.
\newpage
	\newpage
	\section{Introduction}
	The occurrence of liquid droplets on the surface of porous materials and the interaction of liquid water transport through these porous media have applications in a wide range of environmental systems and engineering applications \cite{ackermann2021multi2}. In this study, we focus on the cathode of a polymer electrolyte membrane (PEM) fuel cell. In a PEM fuel cell, hydrogen and oxygen react producing electric energy and water \cite{barbir2012pem}. The produced water can condense into liquid form inside the porous layers in the PEM cathode \cite{andersson2019modeling}. The detailed processes of liquid water evolution in the cathode of a PEM fuel cell are still not well understood \cite{kumbur2009fuel,wang2017effect}. However, the excess water has to be removed from the reaction layer on the cathode side through the porous gas diffusion layer (GDL), a hydrophobic, fibrous material. The GDL shares an interface with the cathode gas distributor which supplies the cell with oxygen (or air) and transports the produced water out of the cell. The cathode gas distributor usually consists of a channel-land structure with straight gas channels \cite{barbir2012pem}. At the interface between GDL and gas distributor channel, liquid water breaks through the porous GDL and forms droplets which grow and are finally detached from the surface and transported away by the surrounding gas flow in the channels. To avoid local flooding, which is a local blockage of the porous GDL by liquid water that prevents the transport of the gaseous reaction components, it is important to analyse the pore-local liquid water transport through the specific GDL structure \cite{kumbur2009fuel,andersson2019modeling,hussaini2009visualization}.\\ Different pore-scale methods have been used in PEM fuel cell applications to simulate the drainage of two-phase flow (water-air) in the cathode gas diffusion layers with different numerical models such as volume of fluid (VoF) \cite{qin2019dynamic,niu2018two,andersson2019modeling,niblett2020two}, Lattice Boltzmann \cite{sakaida2017large,zhang2018three} and pore-network models \cite{straubhaar2015water,aghighi2017pore}. It is found \color{black} that the transport is highly dependent on the liquid water outtake by droplets at the interface between GDL and gas distributor \cite{niblett2020two}. Therefore, it is necessary to include the droplet occurrence, growth and detachment on the two-phase interaction between a porous GDL and the gas distributor channel.\\
	Different approaches to model droplet formation on porous surfaces can be found in the literature. Mularczyk et al. \cite{mularczyk2020droplet} recently performed experimental investigations on the influence of droplet formation at the interface between GDL and gas distributor on the percolation paths of liquid water in the GDL. Santini et al. \cite{santini2013x} considered experimentally the detailed behavior of droplet formation and growth on hydrophobic surfaces. Andersson et al. \cite{andersson2019modeling} used a VoF method to directly describe the droplet behavior on the GDL surface. This method allows to include several physical effects such as drop deformation and contact angle hysteresis. However, the method is computationally quite expensive such that only small computation domains can be considered. Andersson et al. focus on the drop behavior without resolving the GDL structure. In contrast Niblett et al. \cite{niblett2020two} performed a VoF analysis of two-phase flow through a GDL including drop formation but they did not consider drop detachment. Ackermann et al. \cite{ackermann2021multi} developed a more efficient but less accurate formulation to describe drop formation, growth and detachment on the REV-scale. They present a multi-scale approach to couple a porous domain with free flow. The porous domain is discretized on the REV-scale and coupled with a free flow domain through an interface domain, where droplets are described based on their pore-scale behavior. To take the pore-local effects and local displacement processes into account and include their influence on the drop occurrence, we use a pore-network model to describe water flow through the porous GDL structure. Weishaupt et al. \cite{weishaupt2019efficient} developed a coupling approach to combine a pore-network with a free flow model on the pore-scale. At the interface between the domain, they include mass, momentum and energy transfer of the gas phase. In the pore-network, two-phase flow was considered but at the interface only single phase coupling is considered.\\
    In this article, we develop a simplified interface model based on the pore-network principles. In contrast to Ackermann et al. \cite{ackermann2021multi}, we consider the coupling processes between the porous domain and the free flow on the pore-scale. The developed model is an extension to the work of Weishaupt et al. \cite{weishaupt2019efficient} by two-phase coupling processes at the interface. It is now able to describe the droplet behavior on a porous material influenced by a free flow that includes the relevant pore-scale effects but we keep the model efficient with respect to computational costs. With the presented model we bridge the gap between detailed pore-scale methods and efficient but less accurate REV-scale models, which allow the investigation of relevant domain sizes at reasonable computational costs.\\
	The article is structured as follows: First the computation domain of a PEM fuel cell is described. Afterwards, the conceptual and numerical model is presented. The developed model is used to describe both, simplified and more realistic applications. We investigate the behavior of a single drop forming, growing and detaching, the interaction of multiple droplets through the porous medium flow and the droplet formation, growth and detachment at the surface of a porous GDL. The results are compared to numerical and experimental data available in the literature. Finally, the conclusions and a short outlook are presented in the last sections.
	\section{Conceptual and numerical model}
	\subsection{GDL unit cell}\label{sec:GDLunitcell}
	\begin{figure}[h!]
		\centering
		\includegraphics[width=0.8\textwidth]{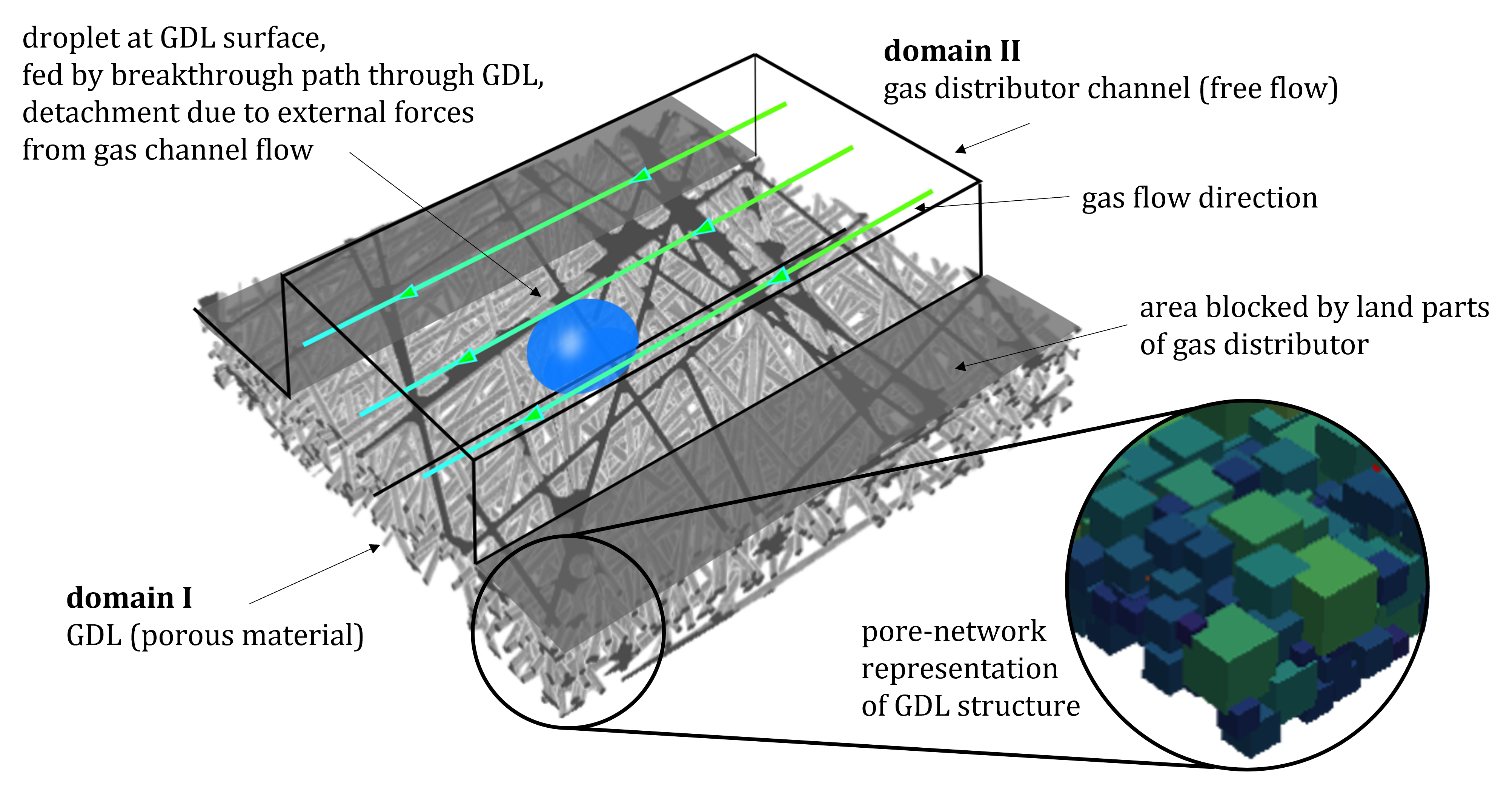}
		\caption{GDL unit cell: $1\times 1 mm$ GDL sample with straight fibers $d=9\mu m$ and porosity $80\%$ in combination with central channel and two half ribs. In the channel, gas flow interactions and droplet formation is considered. In the circle a section of the pore-network representing the GDL structure is shown.}
		\label{fig:GDL-unitCell}
	\end{figure}
	As sketched in Fig. \ref{fig:GDL-unitCell}, simulations are performed in a domain, which is referred to as the GDL unit cell. It corresponds to a section of the cathode GDL located under a central channel and two half ribs. The channel width is $0.5mm$ and the rib half-width is $0.25mm$. The GDL structure is generated using GeoDict \cite{Geodict} with straight fibers of thickness $9\mu m$ and a porosity of $80\%$. The thickness of the GDL is $\delta_{GDL} = 180\mu m$ and we consider a channel height of $H=0.5mm$. To reduce the computation time, a sample of $1\times 1 mm$ is chosen for the investigation. Note that this is not an representative elementary volume but it allows fast calculations and is sufficient to show the working concept of the presented interface drop model applied to a realistic GDL geometry.\\
	The GDL is directly represented by a pore-network which is extracted using the open-source algorithm PoreSpy \cite{gostick2019porespy}. This allows pore-scale investigations corresponding to the real fiber structure of the GDL without specifying pore body and pore-throat size distributions. A section of the pore-network representation is shown in Fig. \ref{fig:GDL-unitCell}. The pore bodies are represented by cubes and the pore throats have a cross-sectional area of the shape of an equilateral triangle to account for corner flow of the wetting phase (gas) and allow contact angles up to $120^\circ$ \cite{blunt2017multiphase}. In the following the GDL is referred to as domain I, while the channel representing the gas distributor as domain II.
	\subsection{Two-phase pore-network model}
	Describing the processes on the pore-scale in domain I (porous GDL), the model needs to be able to describe the following physics:
	\begin{itemize}
		\item two phases $\alpha \in \lbrace \text{gas (wetting) , liquid (non-wetting)}\rbrace$ namely air and water.
		\item immiscible, isothermal two-phase flow (2p)
		\item capillary driven, creeping flow ($Ca \approx 10^{-8}$, $M \approx 50 $, $Re < 1$)
	\end{itemize}
	We extend the fully implicit, dynamic two-phase pore-network model described in \cite{weishaupt2021dynamic}, based on the work of \cite{joekar2012analysis} and \cite{qin2015water}. Joekar-Niasar et al. \cite{joekar2012analysis} developed a dynamic two-phase pore-network model which is accounting for the dynamics of phase displacement processes. Qin \cite{qin2015water} describes an extension of this concept to account for vapor transport. Weishaupt et al. \cite{weishaupt2021dynamic} present a pore-network model coupled to a free flow domain which allows vapor transfer across the interface. We consider two-phase flow without component transport. The new model is able to capture droplet related interface processes and their influence on the dynamic flow processes in the pore-network.\\
	In the following, $s_i^\alpha$ denotes the pore-local saturation of phase $\alpha  \in \left\lbrace \text{gas,liquid}\right\rbrace$ in pore body $i$. $R_i$ is the inscribed radius of a pore body $i$, $r_{ij}$ is the inscribed radius of a pore throat $ij$ and $r_\text{drop}$ is the radius of curvature of a droplet. The subscripts $n$ and $w$ account for the non-wetting and wetting phase, respectively. We consider a hydrophobic pore-network with $n\Leftrightarrow\text{liquid}$ and $w\Leftrightarrow\text{gas}$ . 
	\subsubsection{Mass balance}
	For the 2p pore-network, the continuity equation for a pore body $i$ for each phase $\alpha$ is given by
	\begin{equation}
	V_i \frac{\partial s_i^\alpha}{\partial t} \rho_i^{\alpha} = \sum_j \left(\rho_i^{\alpha} Q^{\alpha}\right)_{ij} + V_iq_i^{\alpha}\,,
	\end{equation}
	with the densities $\rho^{\alpha}$ and a source term for each phase $q^{\alpha}$. The volume of the pore body is denoted by $V_i$.\\
	These equations simplify for incompressible fluids with no local sources and sinks to the volume balance for each phase $\alpha$ in a pore body $i$ with the volume fluxes through the connected pore throats $Q_{ij}$:
	\begin{equation}
	V_i\frac{\Delta s^{\alpha}}{\Delta t} = \sum_j Q_{ij}^{\alpha}\,,
	\end{equation}
	with the change in phase saturation per time step $\displaystyle\frac{\Delta s^{\alpha}}{\Delta t}$.\\
	\subsubsection{Momentum balance}
	Following \cite{weishaupt2021dynamic}, the volume flux $Q_{ij}$ from pore body $i$ to a neighboring pore body $j$ via the pore throat $ij$ is approximated using a Hagen-Poiseuille-type flow for each phase $\alpha$
	\begin{equation}
	Q_{ij}^{\alpha} = k_{ij}^{\alpha}\left(p_i^{\alpha} -p_j^{\alpha} \right)\,.
	\end{equation}
	The phase pressures defined at the pore body centers are given by $p_i^{\alpha}$ and $p_j^{\alpha}$, while $ k_{ij}^{\alpha}$ is a phase-specific pore throat conductance factor which depends on the fluid properties and characteristic geometric features of the pore throats. In pore throats with circular cross-section, the wetting phase vanishes completely during invasion and the wetting phase conductivity is zero after the non-wetting phase invasion. The formulation for the throat conductance factors of the wetting and non-wetting phase used in this study are described in the appendix.
	\subsubsection{Closure relations}
	We use the pore-local capillary pressure saturation ($p_c-s_w$) relation for cubic pore bodies as given by Joekar-Niasar et al. \cite{joekar2012analysis}:
	\begin{equation}
		p_{c,i}(s_{i}^w) = \frac{2\sigma}{R_i \left(1-\exp (-6.83s_{i}^w)\right)}\,,
	\end{equation}
	with the surface tension $\sigma$.\\
	A pore throat can be invaded by the non-wetting phase (here: water), if its capillary entry pressure is exceeded by the local capillary pressure in the connected pore body. The capillary entry pressure for pore throats with a polygonal or circular cross-sectional area is derived from the change in free energy for a displacement of the fluid meniscus in a pore throat following \cite{oren1998extending,blunt2017multiphase}:
	\begin{equation}
		p_{c,e} = \frac{\sigma \cos \theta \left(1 + 2 \sqrt{\pi G}\right)}{r_{ij}} F_d (\theta, G)
		\label{eq:pcEntry}
	\end{equation}
	with the dimensionless function
	\begin{equation}
	F_d(\theta, G) = \frac{1+\sqrt{1+4GD/\cos^2 \theta}}{1+2\sqrt{\pi G}}\,,
	\end{equation}
	and
	\begin{equation}
	G =\frac{A_{ij}}{P_{ij}^2}\,, \qquad
	D = \pi -3\theta + 3 \sin \theta \cos \theta -\frac{\cos^2 \theta}{4G}\,.
	\end{equation}
	Here, $\theta$ is the contact angle, the shape factor is denoted as $G$, $A_{ij}$ is the cross-sectional area of the pore throat between pore body $i$ and $j$, and $P_{ij}$ is the corresponding perimeter of that cross-sectional area.\\
	Snap-off \cite{blunt2017multiphase} occurs if the local capillary pressure falls below
	\begin{equation}
		p_{c,s} = \frac{\sigma\cos(\theta)}{r_{ij}}\left(1-\tan(\theta) \tan(\beta)\right)\,,
	\end{equation}
	with the corner half-angle $\beta = \pi/6$ (for pore throats with a cross-sectional area with the shape of an equilateral triangle).\\
	Finally, we require $s_{i}^w + s_{i}^n = 1$.
	\subsubsection{Primary variables and numerical model}
	We consider a dynamic and fully implicit approach such that both saturations and pressures are solved simultaneously using the Newton-Raphson method. The time step size is chosen adaptive based on the convergence behavior of the non-linear solver (e.g., \cite{baber2012numerical}).\\
	The constitutive relations mentioned above close the system of equations. For the details of the numerical model for the 2p model, we refer to Chen et al., who describe an analogue algorithm to solve dynamic pore-network models in \cite{chen2020fully}.\\
	The model is implemented in the open source framework DuMu$^\text{x}$ \cite{Kochetal2020Dumux}.
	\subsection{Drop formulation in the channel}
\begin{figure}[h!]
		\centering
\begingroup%
  \makeatletter%
  \providecommand\color[2][]{%
    \errmessage{(Inkscape) Color is used for the text in Inkscape, but the package 'color.sty' is not loaded}%
    \renewcommand\color[2][]{}%
  }%
  \providecommand\transparent[1]{%
    \errmessage{(Inkscape) Transparency is used (non-zero) for the text in Inkscape, but the package 'transparent.sty' is not loaded}%
    \renewcommand\transparent[1]{}%
  }%
  \providecommand\rotatebox[2]{#2}%
  \newcommand*\fsize{\dimexpr\f@size pt\relax}%
  \newcommand*\lineheight[1]{\fontsize{\fsize}{#1\fsize}\selectfont}%
  \ifx\svgwidth\undefined%
    \setlength{\unitlength}{150bp}%
    \ifx\svgscale\undefined%
      \relax%
    \else%
      \setlength{\unitlength}{\unitlength * \real{\svgscale}}%
    \fi%
  \else%
    \setlength{\unitlength}{\svgwidth}%
  \fi%
  \global\let\svgwidth\undefined%
  \global\let\svgscale\undefined%
  \makeatother%
  \begin{picture}(1,1.12614538)%
    \lineheight{1}%
    \setlength\tabcolsep{0pt}%
    \put(0,0){\includegraphics[width=\unitlength,page=1]{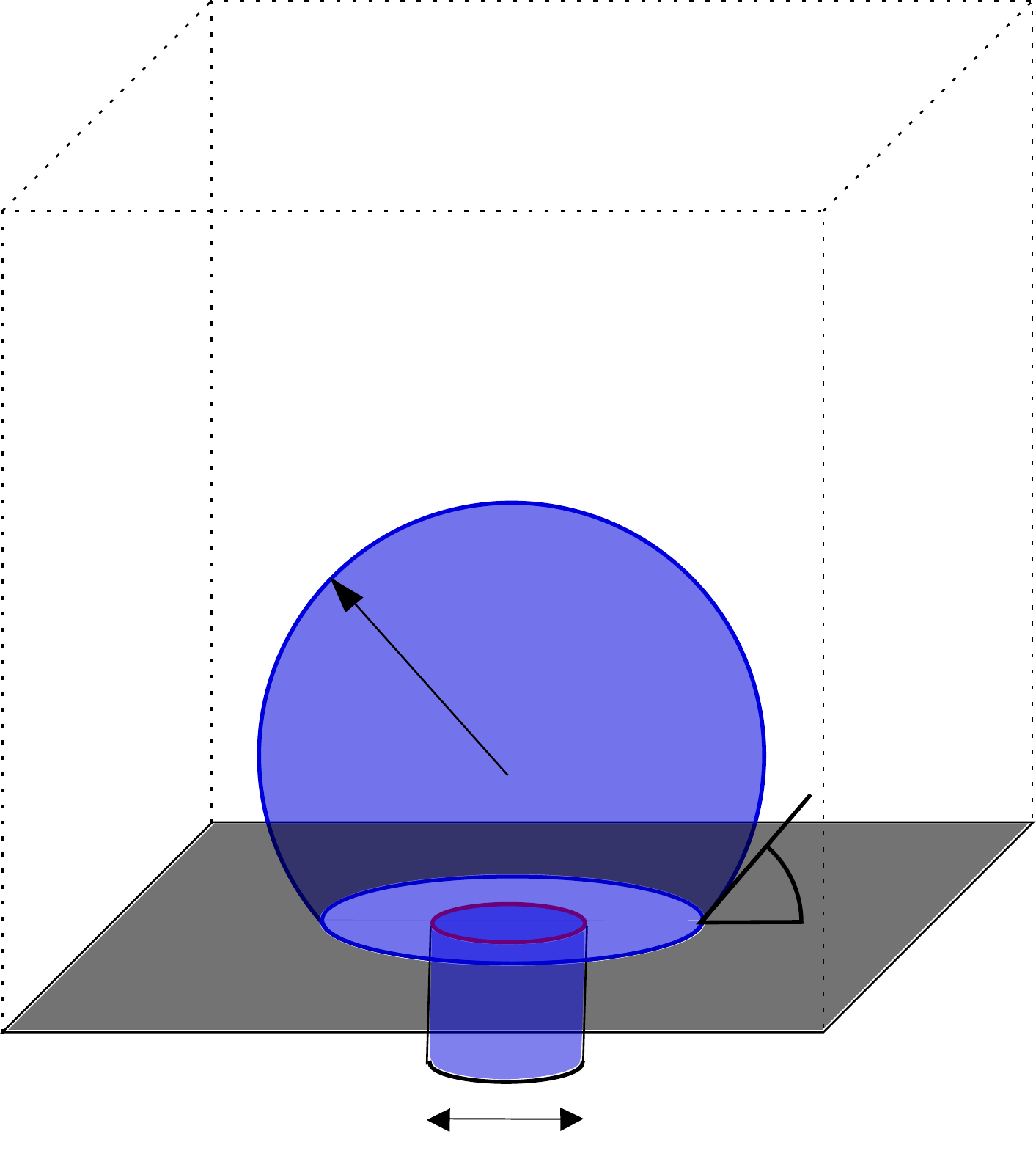}}%
    \put(0.78087469,0.28475562){\makebox(0,0)[lt]{\lineheight{0.62500107}\smash{\begin{tabular}[t]{l}$\theta$\end{tabular}}}}%
    \put(0.41350723,0.48343137){\makebox(0,0)[lt]{\lineheight{0.62500107}\smash{\begin{tabular}[t]{l}$r_\text{drop}$\end{tabular}}}}%
    \put(0.56507495,0.3848347){\makebox(0,0)[lt]{\lineheight{0.62500107}\smash{\begin{tabular}[t]{l}$V_\text{drop}$\end{tabular}}}}%
    \put(0.47689255,0.00549721){\makebox(0,0)[lt]{\lineheight{0.62500107}\smash{\begin{tabular}[t]{l}$r_\text{throat, interface}$\end{tabular}}}}%
    \put(0.47596137,0.79791733){\makebox(0,0)[lt]{\lineheight{0.62500107}\smash{\begin{tabular}[t]{l}$V_\text{pore, interface}$\end{tabular}}}}%
  \end{picture}%
\endgroup%

		\caption{Droplet on a surface in equilibrium without any external forces. The droplet is fed by a liquid water flux through a circular throat. The droplet forms inside an interface pore.}
		\label{fig:DropSurface}
	\end{figure}
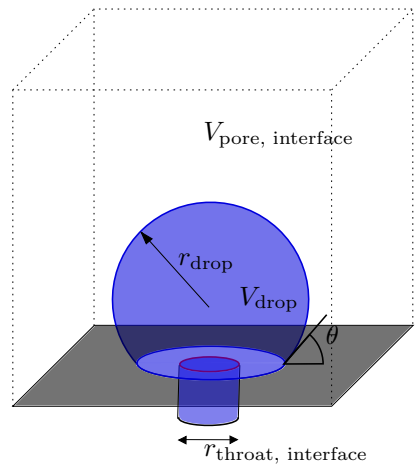

\begin{figure}[h!]
	\centering
	\includegraphics[width=0.5\textwidth]{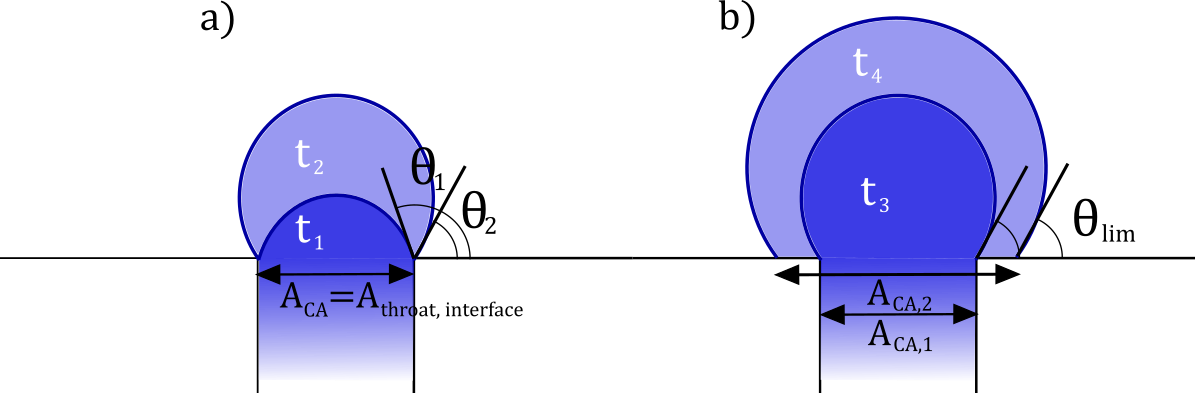}
	\caption{Two stages of drop formation in the interface pores fed by an interface throat.}
	\label{fig:dropStages}
	\end{figure}

\begin{figure}[h!]
		\centering
		\includegraphics[width=0.4\textwidth]{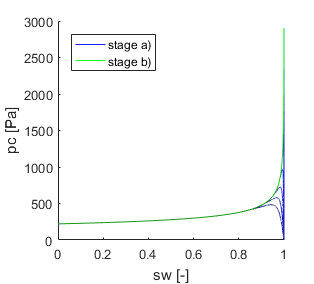}
		\includegraphics[width=0.4\textwidth]{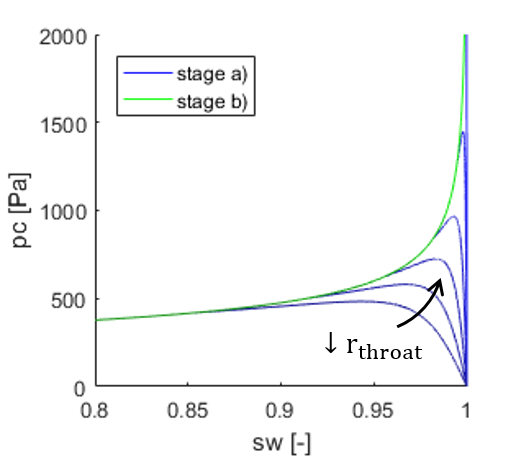}
		\caption{$p_c-s_w$ relation based on geometric analysis describing the drop behavior in interface pores. Left: full wetting phase saturation range, right: zoom in to show the behavior at higher wetting phase saturations (small droplet sizes).}
		\label{fig:pc-sw}
	\end{figure}

	We include the formation of droplets at the interface between domain I and II by the definition of an interface domain. Here, we add interface pores to the throats of the pore-network (domain I) at the border to domain II. In these pores, droplets can form. The behavior of the droplet formation is defined, analogue to the pore-network model, by a local capillary pressure saturation relation. The interface pores are chosen to be large enough to contain the maximum drop volume until the drop detachment. To be consistent with the pore-network model \cite{weishaupt2021dynamic,joekar2012analysis}, we define the contact angle to be measured in the gas phase. \\
	A few assumptions are necessary to describe the droplet in this simplified form:
	\begin{itemize}
	    \item The droplet is small enough such that deformation due to gravity is negligible (Bond number $Bo$ is small enough).
	    \item The deformation of the droplet due to external forces is small enough such that the interface curvature is constant on the droplet surface. This results in a constant pressure inside the droplet.
	\end{itemize}
	Let us consider a droplet on a surface in equilibrium without any external forces. The droplet is fed by a liquid water flux through a circular throat (see Fig.\ref{fig:DropSurface}). The droplet grows inside an interface pore (dotted box in Fig.\ref{fig:DropSurface}), which is independent of the channel domain II but chosen large enough such that the drop can grow in it until the detachment. The detachment criterion is described in Sec.\ref{sec:DropDetachment}.
	The volume of the droplet can be related to the volume of the interface pore, which gives us the saturation in the interface pore:
	\begin{linenomath*}
	\begin{align}
		s_{\text{interface}}^n = 1-s_{\text{interface}}^w= \frac{V_\text{drop}}{V_\text{pore,interface}}\,.
	\end{align}
	\end{linenomath*}
	The curvature of the droplet surface and the corresponding radius of curvature $r_\text{drop}$ depend on the contact angle $\theta$ and the size of the droplet. We get the capillary pressure based on this curvature radius by the Young-Laplace law
	\begin{linenomath*}
	\begin{align}
		p_{c,\text{drop}} = \frac{2\sigma}{r_\text{drop}}\,.
	\end{align} 
	\end{linenomath*}
	To determine the relation between $p_{c,\text{drop}}$ and $s_{\text{interface}}^w$, we distinguish two stages of droplets at the interface:
	\begin{itemize}
		\item stage a) contact area of the drop $A_{CA}$ equals the throat area feeding the drop ($A_{CA} = A_\text{throat, interface}$)
		\item stage b) contact area of the drop $A_{CA}$ is larger than the throat area feeding the drop ($A_{CA} > A_\text{throat, interface}$)
	\end{itemize}
	The two stages are visualized in Fig.\ref{fig:dropStages}.
	\newline
	In stage a), the drop grows with a decreasing contact angle ($\theta_1 > \theta_2$), while the contact area equals the cross-sectional area of the interface throat ($A_{CA} = A_\text{throat, interface}$). This means the contact line is pinned to the corners of the interface throat while the drop is bulging out of this throat. This pinning stays until the critical contact angle $\theta_\text{lim}$ is reached. Here, the critical contact angle equals the static contact angle between the liquid and the solid phase of the porous material. In stage b), the critical contact angle $\theta_\text{lim}$ is reached and stays constant during further drop growing. This results in an increasing contact area ($A_{CA,1} < A_{CA,2}$), while the drop volume increases. With these two stages, for each drop volume (and the resulting saturation in the interface pore), the corresponding drop surface curvature (and the resulting capillary pressure) can be determined based on geometric analysis. The details of this analysis are presented in the appendix. The resulting capillary pressure saturation relation is plotted in Fig. \ref{fig:pc-sw} with the blue part representing stage a) and the green part representing stage b). The relation is dependent on the radius of the feeding interface throat $r_\text{throat, interface}$.
	\newline
	The maximum capillary pressure increases for smaller feeding throats. This results from the smaller sphere cap with a large curvature during the growth in first stage ($t_1$ in Fig.\ref{fig:dropStages}). With larger feeding throats, this peak decreases. The behavior in second stage (stage b)) is rarely influenced by the throat radius. Only the drop volume at which the critical contact angle $\theta_\text{lim}$ is reached shrinks with shrinking throat radius. This extends the curve to higher wetting phase saturations. For low wetting phase saturations ($s_w <0.5$), all curves lay on top of each other. This means, at a certain drop size, the influence of the feeding throat can be neglected.\\
	However, the feeding throat is very important for the drop formation. It determines the capillary entry pressure, that has to be overcome in the pore-network to form a droplet in the interface pore. Therefore, we use the same formulation as for all other throats in the pore-network (Eq.\ref{eq:pcEntry}). The only difference is that the drop-feeding throat has to have circular cross-section in this concept. This avoids corner flow, which would not be defined during stage b) when the contact line is not pinned to the throat radius edge but the drop contact area covers more than just the throat area. Therefore, we choose $G=1/4$ for the feeding throats at the interface.
	\subsubsection{Droplet detachment criterion}
	\label{sec:DropDetachment}
	Since the drop grows at the interface between domain I and II it is not only influenced by the water filling it from domain I but also by the gas flow in domain II (channel). The gas flow results in an external force acting on the droplet, the \textit{drag force} $F_\text{drag}$, while the \textit{retention force} $F_\sigma$ keeps the drop in its position on the surface. If the drag force exceeds the retention force, the droplet is detached from the porous surface. This force balance gives a detachment criterion for droplets forming and growing on a porous surface in a gas flow channel:
	\begin{linenomath*}
	\begin{align}\label{eq:detachmentCriterion}
	    \texttt{if ( } F_\text{drag}(r_\text{drop}) > F_\sigma(r_\text{drop})  \texttt{ ): drop detaches}
	\end{align}
	\end{linenomath*}
	A detailed derivation of the forces is given in the appendix. Here, we just briefly introduce the results.\\
	The drag force acting on the droplet is calculated from a force balance in tangential direction. The derivation results in the following formulation:
	\begin{linenomath*}
    \begin{align*}
    F_\text{drag} &= F_\text{pressure} +F_\text{shear} \\
    &= F_\text{pressure} +F_\text{shear}^\text{top} + 2f_\text{channel}F_\text{shear}^\text{side} \\
    &=  \frac{24 \mu_\text{gas} \left(H/2 \right)^2  \bar{v}_\text{channel} h_\text{drop}^2}{\left(H/2-h_\text{drop}/2\right)^3\left(1-\cos\theta_a\right)^2} + \frac{3\mu_\text{gas} H/2  \bar{v}_\text{channel}}{\left(H/2 -h_\text{drop}/2 \right)^2}\left(2r_\text{drop}\right)^2 + 2 f_\text{channel} \frac{12 \mu_\text{gas} r_\text{drop} h_\text{drop}  \bar{v}_\text{side}}{H}\left(2-\frac{h_\text{drop}}{H}\right) \,,
    \end{align*}
    \end{linenomath*}
    with the channel height $H$ and the height of the droplet $h_\text{drop}$, the fluid viscosity of the gas phase $\mu_\text{gas}$, the average gas flow velocity in the whole channel, $\bar{v}_\text{channel}$, and next to the drop, $\bar{v}_\text{side}$, and the advancing contact angle between the liquid and the solid phase, $\theta_a$.
    This gives a linear relation for the drag force on the average velocity in the channel $\bar{v}_\text{channel}$. For the side shear force, a correction factor $f_\text{channel}$ is applied to account for the reduction of the cross-sectional area of the channel due to the droplet:
    \begin{linenomath*}
	\begin{align}\label{eq:reductionFactor}
		f_\text{channel} = \left(3/2\frac{A_\text{channel}}{A_\text{channel sides}}\right)^2 =  \left(3/2\frac{A_\text{channel}}{A_\text{channel}-2r_\text{drop}H}\right)^2\,,
	\end{align}
	\end{linenomath*}
	dependent on the cross-sectional area of the whole channel $A_\text{channel}$ and the part of the cross-section which is not blocked by the droplet $A_\text{channel sides}$.\\
    This factor is applied to the shear force on both sides of the droplet. A justification of this factor is found in the appendix.\\
    The drop is held in its position by a surface tension force, we call it retention force $F_\sigma$. At the three phase contact line (gas-liquid-solid), a contact angle is formed based on the pair-wise surface tensions between the three phases. In equilibrium state, if no external forces act on the droplet, the static contact angle $\theta$ is formed everywhere on the contact line. External forces cause a deformation of the droplet and a variation of the contact angle along the contact line. This results in an effective force which holds against the external force(s). We follow a formulation for the retention force presented by Kumbur et al. \cite{kumbur2006liquid}:
    \begin{linenomath*}
    \begin{align}
        F_{\sigma} = \sigma\frac{d_{CA}}{2}\pi \left[ \frac{\sin (\Delta \theta - \theta_a) - \sin\theta_a}{\Delta \theta - \pi} + \frac{\sin (\Delta \theta - \theta_a) - \sin\theta_a}{\Delta \theta + \pi} \right]\,.
    \end{align}
    \end{linenomath*}
    with the contact angle hysteresis $\Delta\theta$ (difference between advancing and receding contact angle) and the diameter of the contact area between drop and solid surface $d_{CA}$.\\
    In our case, the retention force acts against the drag force and the drop is first detached when the drag force exceeds the retention force (Eq.\ref{eq:detachmentCriterion}).
	\section{Application to single tube}
	First the formation, growth and detachment of a single droplet is analysed and compared to available data in the literature.
	\subsection{Drop growth without surrounding gas flow}
\begin{figure}[h!]
		\centering
		\includegraphics[width=\textwidth]{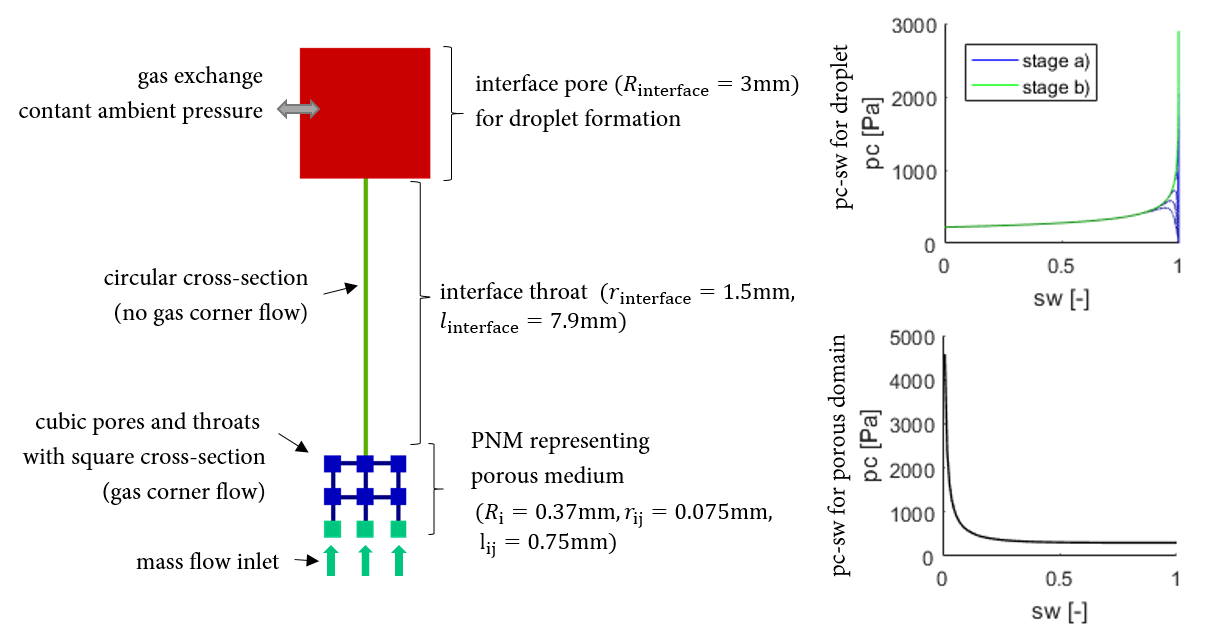}
		\caption{Setup to analyse a single droplet forming in the interface pore of the pore-network model. Green pores: inlet pores ($R_{i} = 3.7\cdot 10^{-4}m$), red pore: interface pore ($R_\text{interface} = 3\cdot 10^{-3}m$), green throat: drop-feeding throat ($r_\text{interface}=1.5\cdot 10^{-3}m$), dark blue throats: network throats ($r_{ij} = 7.5\cdot 10^{-5}m$). On the right side, the local capillary pressure saturation relations are visualized. Top: interface pore with two stages of drop growth, bottom: regular network pore in hydrophobic porous medium.}
		\label{fig:SetupSingleDrop}
	\end{figure}

\begin{figure}[h!]
		\centering
		\includegraphics[width=0.6\textwidth]{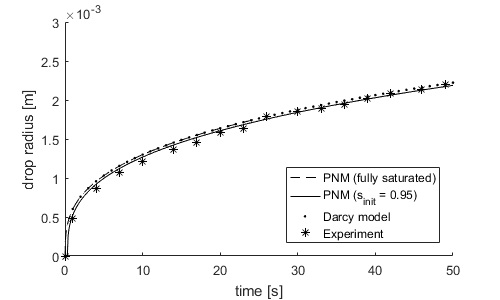}
		\caption{Drop radii from pnm simulations with different initial saturations (fully saturated: $s_\text{init}^n = 1.0$,  $s_\text{init}^n=0.95$) and Darcy simulation calculated by Ackermann et al. \cite{ackermann2021multi}}
		\label{fig:DropRadiusBergamo}
	\end{figure}

	In this section, we compare the drop growth behavior of the developed model to experimental data and a comparable drop model previously presented in the literature. Both data sets are presented by Ackermann et al. in \cite{ackermann2021multi}. We analyse the droplet behavior without any surrounding gas flow. Similar to the REV-scale model presented in \cite{ackermann2021multi}, a very simple setup is used to analyse the growth of a single droplet (see Fig.\ref{fig:SetupSingleDrop}). We consider a simple 3D pore-network.
	\newline
	On the right side in Fig.\ref{fig:SetupSingleDrop}, the local capillary pressure saturation relations of the pores in the network are visualized. The curve on the top shows the relation corresponding to the interface pore (red) with the two stages of drop growth. The shape depends on the radius of the interface throat (green). The relation shown at the bottom of the figure represents the local capillary pressure saturation relation of a regular network pore in hydrophobic porous medium. This relation is applied to the network at the bottom of the single tube setup (blue and green pores).\\
	The model setup is designed as a mixture of the Darcy framework in \cite{ackermann2021multi} and the experimental setup described in \cite{ackermann2021multi}. The feeding throat for the interface pore equals the tube used in the experiment and a pore-network is applied to represent the porous domain of the Darcy model, where the inlet condition is applied. For the pore-network, a completely homogeneous two-dimensional network is chosen with pore body inscribed radii of $R_i= 3.7\cdot 10^{-4}m$, pore throat inscribed radii $r_{ij} = 7.5\cdot 10^{-5}m$ and pore throat length $l_{ij} = 7.5\cdot 10^{-4}m$. The feeding tube in the experiment has a length of $7.9mm$ and a radius of $1.5mm$. Therefore, the feeding throat in the pore-network model is chosen accordingly ($l_\text{interface}=7.9\cdot 10^{-3}m$, $r_\text{interface}=1.5\cdot 10^{-3}m$). The interface pore is large enough to capture a droplet up to $2.5mm$ as it is expected from the experiment and Darcy model ($R_\text{interface} = 3\cdot 10^{-3}m$).
	\newline
	The comparison of the droplet growth considering the drop radius in time shows very good agreement between the pore-network model (dashed and solid lines in Fig.\ref{fig:DropRadiusBergamo}) and the experimental data (stars in Fig.\ref{fig:DropRadiusBergamo}). A change in the initial saturation of the pore-network result in a time shifting of the curve. With a lower initial liquid saturation ($s_\text{init}^n=0.95$), the drop formation starts a little later. This results from the capillary entry pressure which has to be overcome to invade the interface throat which feeds the droplet. Therefore, the pore-network is first filled with liquid before the droplet starts forming. A perfect match of the numerical data (in Fig.\ref{fig:DropRadiusBergamo}: Darcy model dots, pore-network model dashed and solid lines) is achieved, using an initial water saturation $s_\text{init}^n=1.0$ (fully saturated).
	\subsection{Drop detachment from a single tube by drag force from surrounding gas flow}
\begin{figure}[h!]
		\centering
		\includegraphics[width=0.8\textwidth]{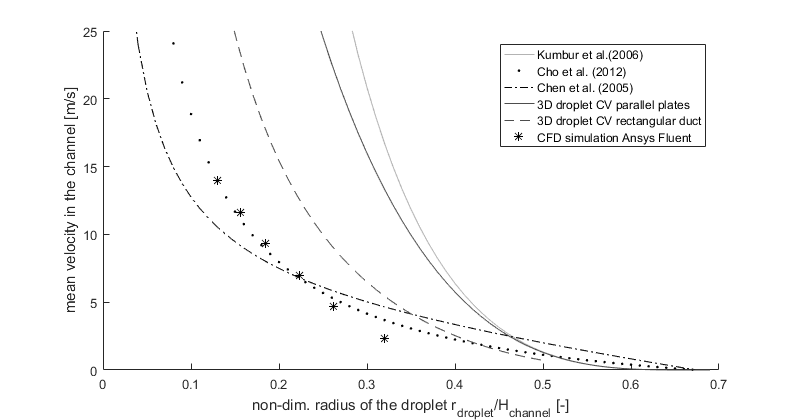}
		\caption{Detachment lines from different derivation approaches. Minimal average velocity for detachement with respect to non-dimensionalized drop radius (non-dimensionalized with respect to the channel height $H_{channel}$). Solid lines: completely analytical derivation, stars: numerical calculation, dashed/dotted lines: empirical and mixed approaches.}
		\label{fig:separationLines}
	\end{figure}

	In this section, we compare the developed drop detachment criterion with available concepts from the literature and with a CFD volume of fluid (VoF) simulation of a droplet emerging from a single tube in a channel. A flow velocity is applied to the upstream end of the channel such that a drag force from a parabolic gas flow profile acts on the droplet causing its deformation and finally detachment. We use the commercial software ANSYS Fluent for the investigations. The details of the VoF model setup are described in the appendix.\\
	We investigate the droplet radius based on the volume of the detached droplet. Therefore, we consider the drop as a sphere segment. The radius and volume based on the contact area $CA$ and the contact angle $\theta$ are given by: 
	\begin{linenomath*}
	\begin{align}
	    r_\text{drop}&= \frac{h_\text{drop}}{1-\cos{\theta}} = \frac{d_{CA}}{2\sin\theta} \\
     V_\text{drop} &= \frac{\pi}{3}h_\text{drop}^2(3r_\text{drop}-h_\text{drop}) = \frac{\pi}{3}\left(\frac{d_{CA}}{2}\right)^3\frac{(1-\cos(\theta))^2(2+\cos(\theta))}
     {(\sin(\theta))^3}\,,
	\end{align}
	\end{linenomath*}
    where $r_\text{drop}$ is the radius of the drop curvature, $h_\text{drop}$ is the drop height, $V_\text{drop}$ is the volume corresponding to the sphere cap shaped (undeformed) droplet and $d_{CA}$ is the diameter of the contact area between liquid and solid. From the VoF simulation, we measure the drop volume and calculate the corresponding drop radius for the comparison.\\
    Since we are interested in an appropriate detachment criterion, we analyse the \textit{separation line} which gives the critical mean velocity at which a droplet with a certain radius will be detached. In Fig.\ref{fig:separationLines}, six different approaches to determine the critical velocity for drop detachment are visualized.
	Kumbur et al. \cite{kumbur2006liquid} expound a two-dimensional (2D) description of the force balance around a droplet on a solid surface which is influenced by surrounding gas flow. In the retention force, the authors include the contact angle hysteresis, and in the drag force, the pressure and shear forces at the top of the droplet are taken into account. The resulting detachment or separation line is based on the balance of these two forces. The droplet is detached once the drag force exceeds the retention force. The 2D approximation represents an infinitely wide channel (parallel plates) but also an infinitely wide droplet (deformed cylindrical shape) such that shear forces at the sides of the droplet are not included. The model presented by Kumbur et al. \cite{kumbur2006liquid} is derived completely analytical. In contrast, Chen et al. \cite{chen2005simplified} developed a semi-analytical approach by using an experimental fitting for the drag coefficient $c_D$. The used experiment for the fitting represents an ex-situ PEM fuel cell cathode setup. For the retention force, they use a similar approach as Kumbur et al. \cite{kumbur2006liquid}. Cho et al. \cite{cho2012droplet} extended the concept presented in \cite{chen2005simplified} by further developing the contact angle formulation along the contact line. For the drag force Cho et al. \cite{cho2012droplet} use the same drag coefficient formulation as Chen et al. \cite{chen2005simplified}. These empirical approaches are plotted with dotted and dashed-dotted lines in Fig.\ref{fig:separationLines} while completely analytical approaches are visualized with solid lines. The black line (3D droplet CV parallel plates) shows an extension of the analytic model of Kumbur et al. \cite{kumbur2006liquid}. Kumbur et al. \cite{kumbur2006liquid} considers a two-dimensional model representing an infinite wide channel and a cylindrical shape droplet. The new 3D model (3D droplet CV parallel plates) considers a spherical droplet in an infinite wide channel, where the shear forces at the sides of the spherical droplet are included. For the derivation of the separation line, a control volume (CV) is placed around the droplet in the channel. At the top surface of the CV the top shear force is calculated similar to the approach presented by Kumbur et al. \cite{kumbur2006liquid}. Additionally, the shear forces at the sides of the droplets are taken into account which allows the consideration of a spherical drop in the wide channel. In the concept resulting in the dashed, black line (3D droplet CV rectangular ducts), the reduction of the cross-sectional area due to the droplet is considered. This is done using the reduction factor in Eq.\eqref{eq:reductionFactor} and  allows arbitrary aspect ratios of the channel. The detailed derivations of the CV approaches are presented in the appendix.\\
    The asterisks show the detachment evaluations of six different VoF simulations with the corresponding mean gas flow velocity resulting from an inlet condition with parabolic flow profile. The lowest mean velocity analyzed is $\text{v}_\text{mean}= 2.32m/s$ (corresponding $\text{v}_\text{max}= 5m/s$). At lower gas flow velocities, no detachment due to the gas flow drag force occurred but the droplet touched the channel walls which are chosen hydrophilic. Higher gas flow velocities than $\text{v}_\text{mean}= 13.94m/s$ (corresponding $\text{v}_\text{max}= 30m/s$) have not been analysed. Especially for low gas flow velocities, the VoF simulation results differ from the other data resulting in smaller drop sizes at detachment.\\
	Overall, the data lines in Fig.\ref{fig:separationLines} widely spread. For the analytic derivations inaccuracies result from the chosen simplifications while for the empirical approaches the uncertainty of the data is more difficult to determine.\\
	Further investigations are necessary to determine the "correct" separation line and reduce the uncertainties.\\
	For the pore-network drop model, we choose the analytic CV approach derived for three-dimensional rectangular channels (3D droplet CV rectangular duct). The approach is based on a force balance (see Sec.\ref{sec:DropDetachment}) and clear assumptions and simplifications which can be adapted, if further knowledge on the system is available.
	\section{Application to multi-tube system}
	In this section, the ability of the developed model to simulate the formation, growth and detachment of several droplets at the interface between a porous domain and a free-flow domain is shown and compared to experimental findings from the literature.
	\subsection{Comparison with experimental data}
\begin{figure}[h!]
		\centering
		\includegraphics[width=0.5\textwidth]{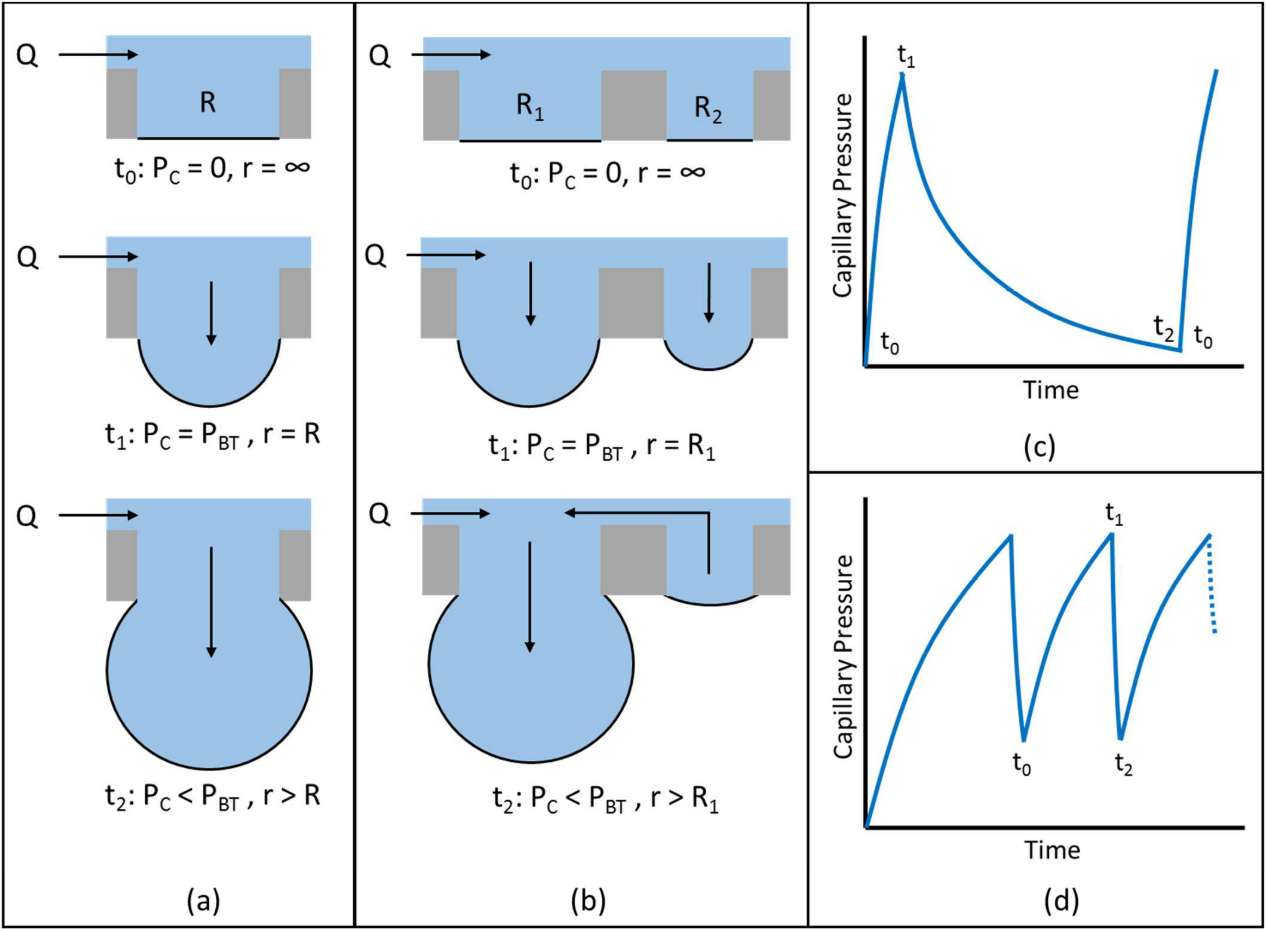}
		\caption{Bulging menisci model: Schematic of pressure profiles observed during droplet grow-detachment cycle. (a) Single tube, (b) Multiple tubes with a single breakthrough site, (c) Pressure profile for the single tube, (d) Pressure profile for the multi-tube system \cite{quesnel2015dynamic}.}
		\label{fig:QuesnelModel}
	\end{figure}

\begin{figure}
		\centering
		\includegraphics[width=\textwidth]{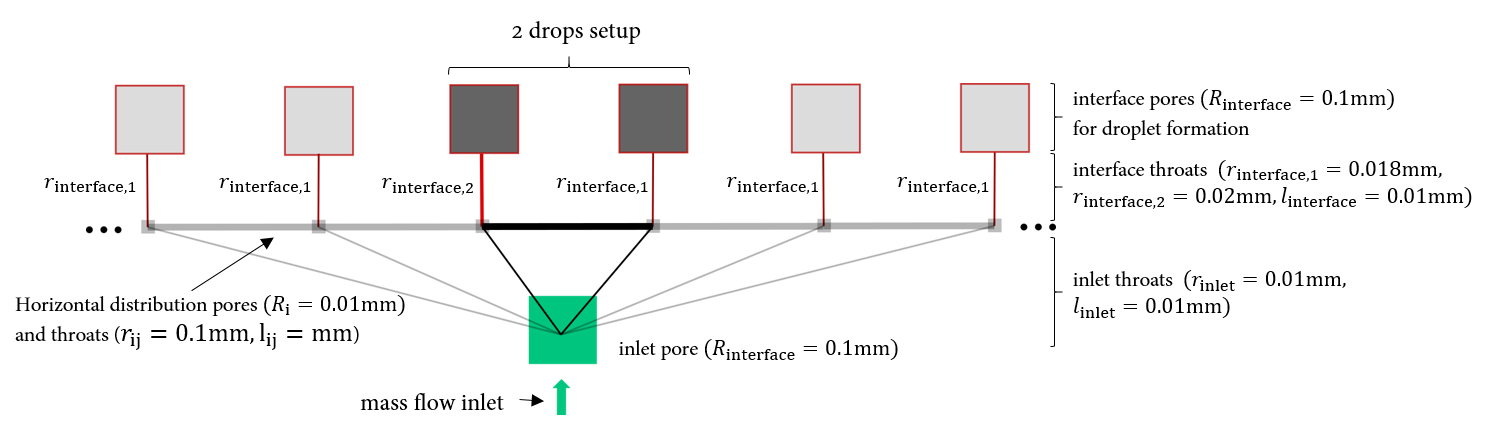}
		\caption{Geometric pore-network configurations to analyse the interactive behavior of several drops in a multi-tube system. Different configurations have been analysed: 2 pores, 16 pores, 101 pores.}
		\label{fig:MultiTubeSetups}
	\end{figure}

\begin{figure}[h!]
		\centering
		\includegraphics[width=0.45\textwidth]{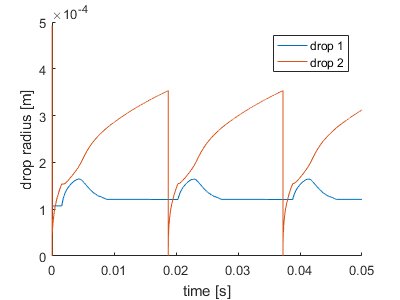}
		\includegraphics[width=0.45\textwidth]{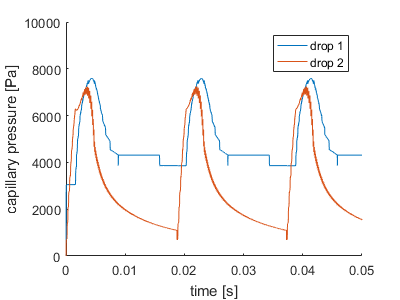}
		\caption{Two-drop interaction. Left: drop radius evolution with time of the droplets forming at the two different interface throats. Right: Corresponding capillary pressure of the droplets (capillary pressure is defined by the interface curvature of the drop surface).}
		\label{fig:TwoDropInteraction}
	\end{figure}

\begin{figure}[h!]
		\centering
		\includegraphics[width=0.7\textwidth]{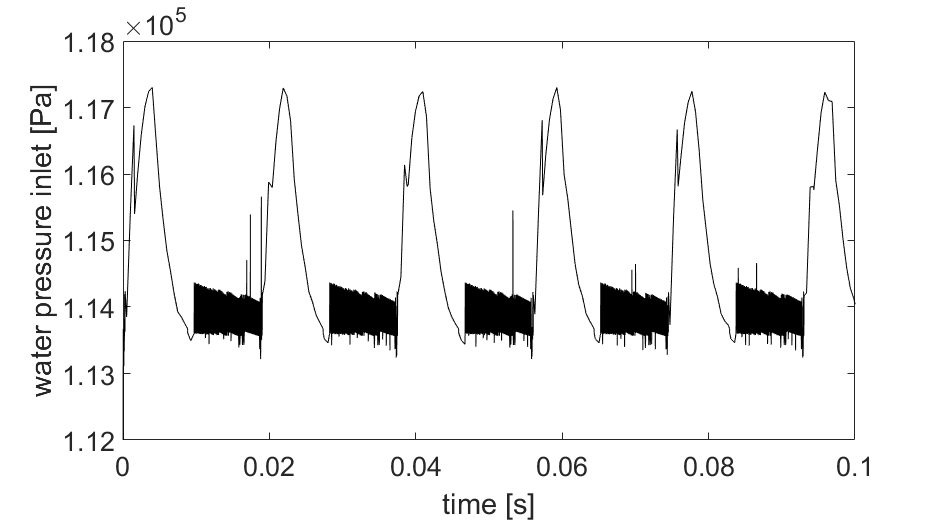}
		\caption{Water pressure in the inlet pore of a multi-tube setup with two interface throats to fill droplets and two connected interface pores for droplets to grow in.}
		\label{fig:p_liq_inlet-Gostick3}
	\end{figure}

\begin{figure}[h!]
		\centering
		\includegraphics[width=0.45\textwidth]{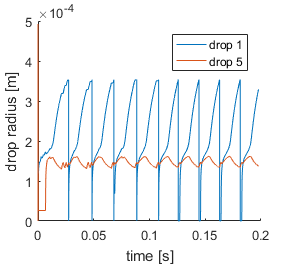}
		\includegraphics[width=0.45\textwidth]{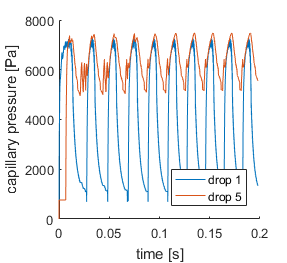}
		\caption{Two-drop interaction. Left: drop radius evolution with time of the droplets forming at the two different interface throats. Right: Corresponding capillary pressure of the droplets (capillary pressure is defined by the interface curvature of the drop surface).}
		\label{fig:TwoDropInteraction16}
	\end{figure}

\begin{figure}[h!]
		\centering
		\includegraphics[width=0.45\textwidth]{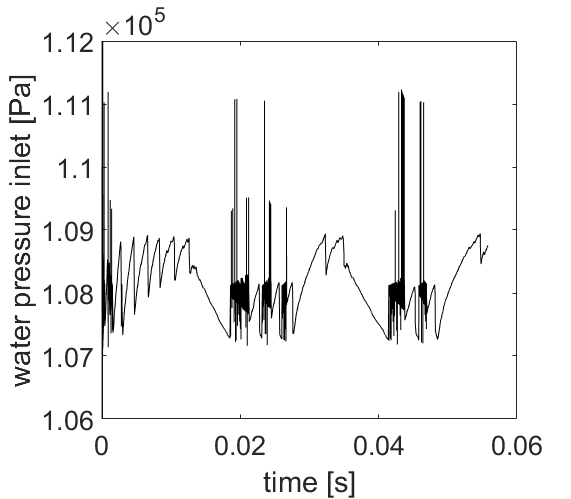}
		\includegraphics[width=0.45\textwidth]{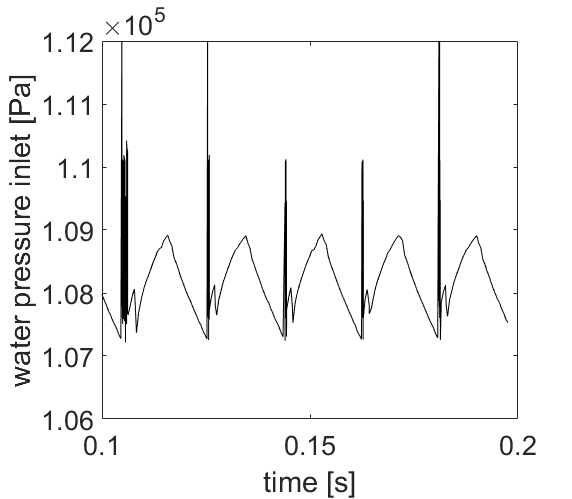}
		\caption{Liquid phase pressure in the inlet pore of the 16 interface pores setup. Left: first three drop formation cycles. Some numerical fluctuations occur. But the system stabilizes and less fluctuations occur at later time (right). Here, a repeating behavior of the pressure curve establishes for drop formation, growth and detachment.}
		\label{fig:p_liq_inlet-Gostick16}
	\end{figure}

\begin{figure}[h!]
	\centering
	\includegraphics[width=0.7\textwidth]{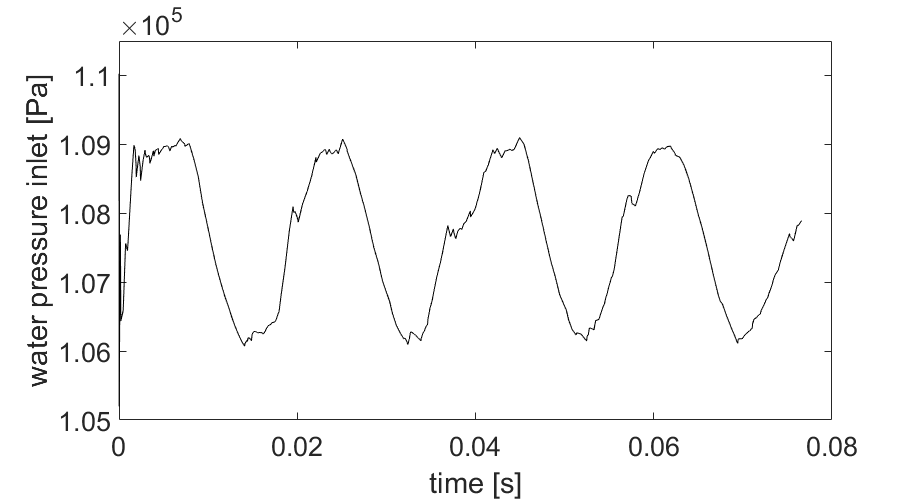}
	\caption{Liquid phase pressure in the inlet pore of the 101 interface pores setup.}
	\label{fig:p_liq_inlet-Gostick101}
\end{figure}

	Quesnel et al. present experimental investigations of multi drop interactions in their publication \cite{quesnel2015dynamic}. They explain the growth behavior of several droplets forming at an interface from feeding throats of different size. Due to the capillary pressure changes in the droplet during the growth, the initially faster growing droplets are favoured and are fed by the surrounding smaller droplets which decrease as a result. This is caused by the non-monotonic capillary pressure curve. \\
	They also present the so called \textit{bulging menisci model} which is similar to a common \textit{bundle-of-tubes model}, but the volume of fluid in the tubes themselves only considers the inflation and deflation of menisci with time \cite{quesnel2015dynamic}. In Fig.\ref{fig:QuesnelModel}, the interactive behavior of the forming droplets is visualized. For details on the model, it is referred to the developers \cite{quesnel2015dynamic}.
	\newline
	The bulging menisci model explains the behavior of the drops forming on a GDL surface by simplifying the geometrical setup to straight tubes rather than the complex fiber structure.\\
	In this section, we show that the interactive behavior of multiple droplets in the system is automatically included in the developed model. Therefore, we use a simplified geometry and consider a setup with 2, 16 and 101 tubes (see Fig.\ref{fig:MultiTubeSetups}).
	\newline
	For the two-drop interaction, we consider two interface pores which are connected by an interface throat to the network. The two interface throats have a different size such that one of them will be favored during the drop growth caused by the flow resistance in the throat ($r_\text{interface,1}= 2\cdot 10^{-5}m$ and $r_\text{interface,2}= 1.8\cdot 10^{-5}m,$ ). The network pores are much smaller than the interface pores ($R_\text{network} = 5\cdot 10^{-5}m$ and $R_\text{interface} = 5\cdot10^{-4}m$) to prevent large amounts of water storage in the network. The network pores are connected with wide, short throats ($r_{ij,\text{horizontal}} = 3\cdot 10^{-5} m, l_{ij,\text{horizontal}} = 10^{-5}m$) to ensure a fast exchange (large flux) between the forming droplets. As inlet condition, a water mass flux is applied to the inlet pore. The gas phase pressure in the interface pores is kept constant by a Dirichlet boundary condition.\\
	For the setup with 16 interface pores, not only two but 16 interface pores are placed next to each other. Again, one interface throat is larger than the other ones to be favored during the drop growth. For the setup with 101 interface pores, ten interface throats are chosen to be larger than the rest. Since inertial forces are negligible for the creeping flow in the network, the position of the interface pores is not important.\\
	In Fig.\ref{fig:TwoDropInteraction}, the drop radius and corresponding local capillary pressure in the interface pores are shown for the two-interface-pores investigation. It can be seen how both droplets start growing but then one of them is favored while the other drop volume is decreasing. The favored droplet growing rate increases with the shrinkage of the other one since the volume of drop 1 is transferred to drop 2. At a certain size, the larger droplet is detached (drop raidus drops to zero). Here, this is due to surrounding gas flow which results in a drag force acting on the droplet. The local capillary pressure curves in Fig.\ref{fig:TwoDropInteraction} explain why drop two is favored. The larger drop volume results in a smaller curvature and therefore a lower pressure in the droplet. The gas phase pressure is kept constant such that the shape of the curves equals the one of the water pressure in the droplet.
	In Fig.\ref{fig:p_liq_inlet-Gostick3}, the water pressure in the inlet pore of the two-interface-pores setup is shown. The shape is similar to the capillary pressure in the interface pores. The fluctuations at the lower level, when the drop volume of the favored drop increases while the other one stays constant, result from numerical instabilities in the small dynamic network.
	\newline
	With a constant gas phase pressure in the drop pores, the shape of the curve equals the global capillary pressure ($P_{c,\text{global}} = p_\text{liq, inlet}-p_\text{gas, drop}$).\\
	Therefore, the interaction of only two droplets in the simple network is not sufficient to explain the change in the global capillary pressure behavior in Fig.\ref{fig:QuesnelModel}. In the experiments in \cite{quesnel2015dynamic}, a GDL structure is used where more than two droplet occurred and interacted. In the next step, a network with 16 parallel interface pores is analysed. Again, one interface pore is chosen to be favored by connecting it with a larger interface throat. In Fig.\ref{fig:TwoDropInteraction16}, two droplets in the network with 16 interface pores are compared. The drop radius shows a similar behavior as in the two-interface-pores setup. However, the growing rate before and after the shrinkage of the other droplets differs more than before. This results from the larger number of droplets which are formed and where the water mass flux is distributed to and the resulting larger number of shrinking droplets and therefore the larger volume which is transferred to the favored droplet afterwards. The local capillary pressures in the interface pores look similar than before but with a few numerical fluctuations.
	\newline
	In Fig.\ref{fig:p_liq_inlet-Gostick16}, the corresponding water phase pressure in the inlet pore for the setup with 16 interface pores is shown. Here, a clear difference can be seen to the behavior of the water pressure in Fig.\ref{fig:p_liq_inlet-Gostick3}. The fast increase of the pressure is hindered such that it almost equals the rate of pressure decrease. Again, we see some numerical fluctuations in the pressure during the growth of the favored droplet and while the size of the other droplets stays nearly constant (except small numerical fluctuations). The overall behavior now gets closer to the pressure behavior which has been predicted by the bulging menisci model.
	\newline
	To further analyse the behavior of the water pressure in the inlet pore and therefore the global capillary pressure during the multi-drop interaction, a setup with 101 interface pores is analysed. In this case, ten interface throats are slightly larger than the rest such that droplet formation will be favored. The resulting water pressure behavior in the inlet pore is presented in Fig.\ref{fig:p_liq_inlet-Gostick101}.
    \newline
    In this simulation, the average rate of pressure growing is smaller than the one of pressure decreasing. The behavior is not as strong as expected from the bulging menisci model but a trend can be seen from the analysed simple test cases with 2, 16 and 101 interface pores. The trend shows, the more droplets strongly interact with each other the closer the water pressure behavior is to the predicted curve. For a large GDL sample, where many droplets occur and interact, a long pressure increase followed by an almost sudden decrease might occur.
	\section{Application to GDL sample}
\begin{figure}[h!]
		\centering
		\includegraphics[width=0.5\textwidth]{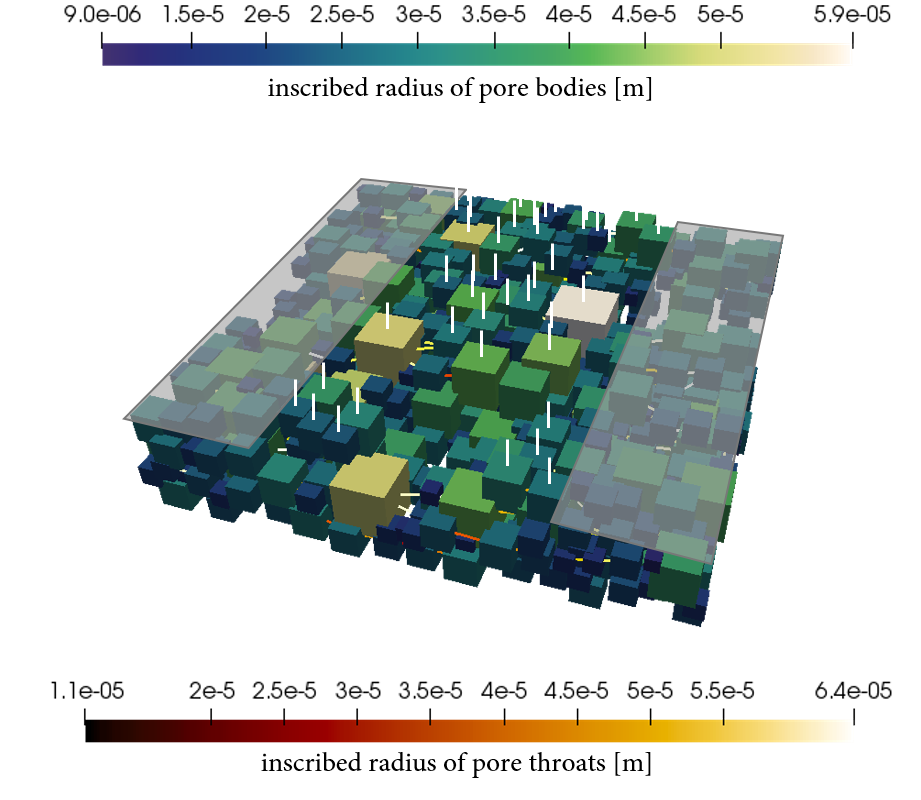}
		\caption{Pore-network representing a GDL structure. The white lines mark the locations of open pores in the channel, where droplets can form. The gray shaded areas are part of the ribs. Here, the network pores are blocked and have no access to the channel.}
		\label{fig:PN-GDL-dropPaper}
	\end{figure}

\begin{figure}[h!]
	    \centering
	    \includegraphics[width=0.45\textwidth]{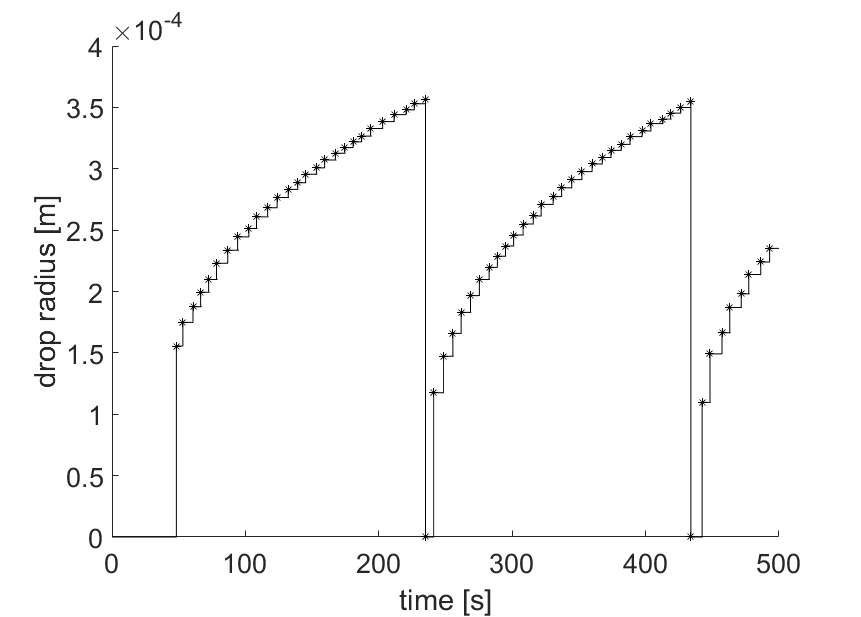}
	    \includegraphics[width=0.45\textwidth]{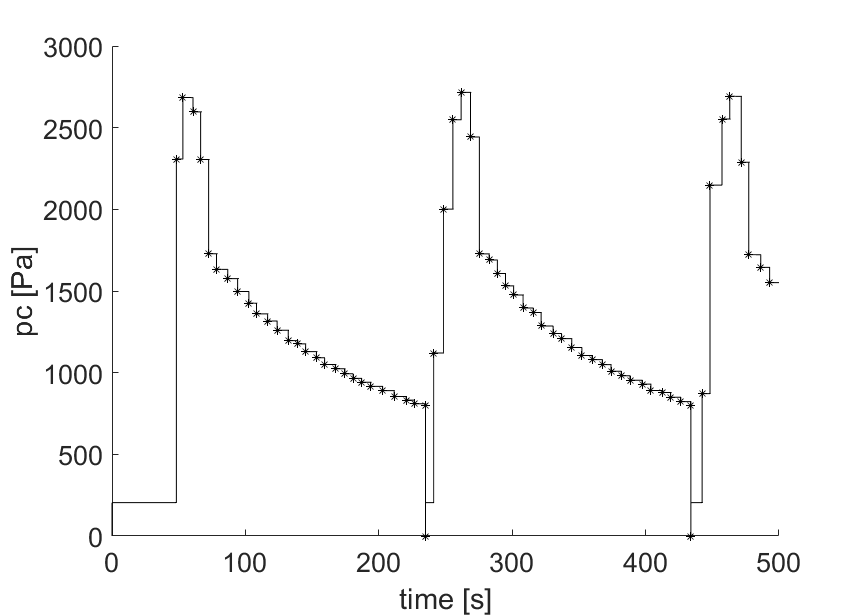}
	    \caption{Left: droplet radius, right: corresponding capillary pressure in the interface pore (capillary pressure at droplet surface)}
	    \label{fig:DropGDL}
	\end{figure}

	In this section, the pore-network model including the interface concept to model drop occurrence, growth and detachment is applied to a network representing a small GDL unit cell as presented in Sec.\ref{sec:GDLunitcell}. In Fig.\ref{fig:PN-GDL-dropPaper}, the pore-network representing a GDL structure is visualized. The network resulting from the fiber structure shows wide pore-size and throat-size ranges. We apply a mass flow boundary condition to the bottom side. Therefore, an inlet pore is added which is connected to each pore at the bottom side of the network, which represents the water production rate corresponding to a current density of $1\,$A$/$cm$^2$. The connecting throats equal the size of the network pores such that the mass flux is distributed to all inlet pores based on their pore size due to the capillary entry pressure of the connecting throats.
	\newline
	The considered GDL sample is small enough such that only one liquid water breakthrough location occurs where a droplet forms, grows and finally detaches. The drop radius and the corresponding local capillary pressure are shown in Fig.\ref{fig:DropGDL}. The step wise behavior results from the low water inlet flux and small time step sizes. Locally in the network, the capillary entry is overcome in a pore but the flux through the freshly invaded throat results in a decrease of the local saturation and therefore a decrease of the local pressures. These anomalies based on the pore-network model nature result in step wise fluxes in the network.
	\subsection{Influence of the drop formation at the interface between GDL and gas distributor on the flow in the GDL}
\begin{figure}[h!]
		\centering
		\includegraphics[width=\textwidth]{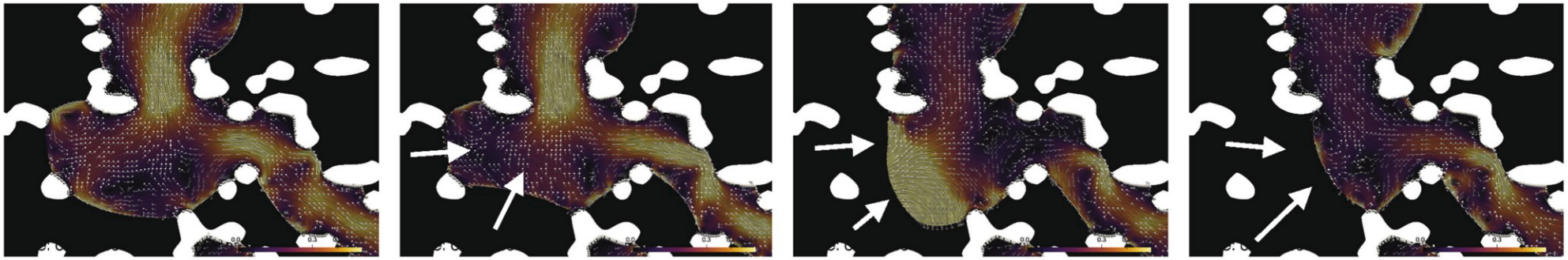}
		\caption{Close up of dynamic receding of pore contents in the analysed GDL structure resulting from the growth of droplet on the surface of the GDL in the GC (gas channel). Velocity magnitude and vectors represent direction of flow \cite{niblett2020two}.}
		\label{fig:NiblettReceeding}
	\end{figure}

\begin{figure}[h!]
	    \centering
	    \includegraphics[width=0.9\textwidth]{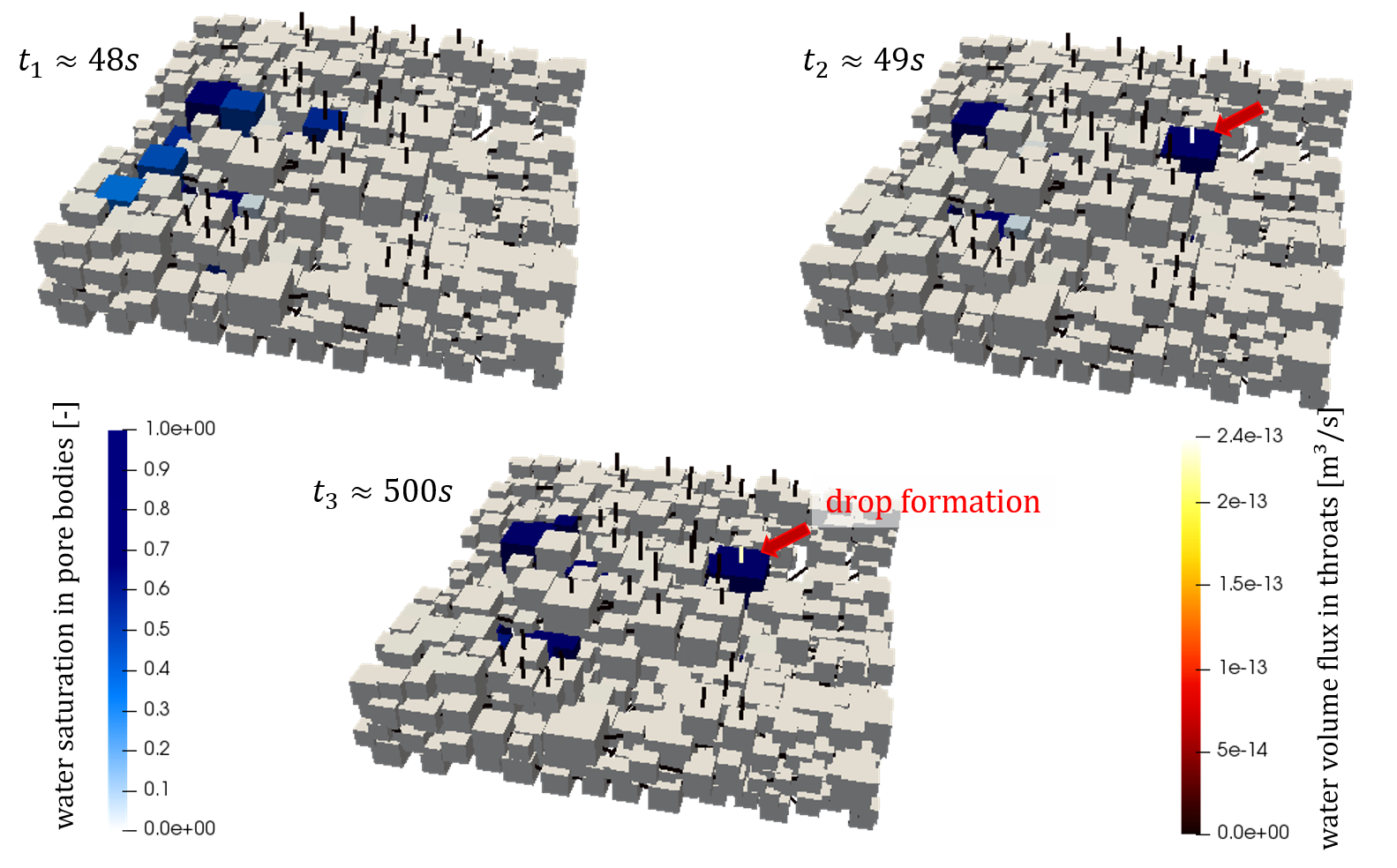}
	    \caption{Top left: local water saturation pattern in the pore bodies and water volume flux in the throats before breakthrough and drop formation ($t\approx 48$s), top right:  local water saturation pattern in the pore bodies and water volume flux in the throats after breakthrough ($t\approx 49$s), the water flux through the interface throat forms a droplet in the interface pore. Bottom: established local water saturation pattern in the pore bodies and water volume flux in the throats after multiple drop formations ($t\approx 500$s). The final pattern shows the stabilized breakthrough path.}
	    \label{fig:saturationGDL}
	\end{figure}

	Not only the droplets forming at the interface betwen GDL and gas distributor are influenced by the water displacement and fluxes in the pore-network but also the pore-network itself is influenced by the formation, growth and detachment of the droplets. The developed model is capable to capture these influences pore-locally. In the following section, the developed model is again applied to the previously described unit cell and the flow behavior inside the GDL is analysed during the drop formation, growth and detachment. The behavior during drop formation and growth is compared to computational fluid dynamics (CFD) investigations presented by Niblett et al. \cite{niblett2020two}.\\
	Niblett et al. \cite{niblett2020two} present a CFD model (VoF) that allows analysis of two-phase flow in a GDL structure. The results show a highly dynamic flow behavior. Especially during breakthrough (and the formation of droplets), when menisci recede from certain pores (see Fig.\ref{fig:NiblettReceeding}).
	In Fig.\ref{fig:NiblettReceeding}, a VoF simulation of two-phase flow through the two-dimensional cross-section of a GDL is shown. The visualization focuses on a single breakthrough path of water in the void space (black) between fibers (white). The breakthrough path feeds a droplet at the GDL surface (at the top of the figures). The growing droplet is not shown in the figure. In the water phase, the flow velocity is visualized. Light colored areas indicate a higher velocity than the darker ones. The four presented time steps in Fig.\ref{fig:NiblettReceeding} present the behavior of a certain meniscus, which is receding during the formation of a droplet fed by this breakthrough path. In Fig.\ref{fig:NiblettReceeding}, white arrows point at the receding meniscus. A larger droplet has a lower water phase pressure such that the water tends to fill the droplet rather than pores in the porous medium. This results in receding menisci in certain pores.\\
	This behavior is also captured by the pore-network model. In Fig.\ref{fig:saturationGDL}, the local saturation at three different time steps are presented. In the pore throats, the water volume flux is shown, which is zero in most of the visible throats since the water flows through the preferential flow paths. At $t\approx 48s$, the water reaches the pores at the top side of the GDL but breakthrough did not occur yet. Several pores are visibly invaded and filled with water but all of the interface throats are black which indicates zero water volume flux. At $t\approx 49s$, breakthrough occurred at one of the pores located in the channel region of domain II (marked with the red arrow). A flux goes through the interface throat filling a droplet in the interface pore (the interface pore is not visible in Fig.\ref{fig:saturationGDL}). At the same time, the saturation in several pores, connected by invaded throats, decreased. This behavior equals the receding menisci observed in the CFD analysis by Niblett et al. \cite{niblett2020two} (Fig.\ref{fig:NiblettReceeding}). Since the developed pore-network model allows the simulation of dynamic droplet formation and detachment, the behavior of the network during several growth and detachment periods can be analysed. It shows, for the small GDL unit cell with only one breakthrough, a stable breakthrough path establishes ($t\approx 500s$ in Fig.\ref{fig:saturationGDL}), which does not change significantly during the simulation. The flux through the interface throat into the interface pore, where the droplet is forming is slightly decreased compared to $t\approx 49s$.
	\section{Discussion and outlook}
	The developed model is able to simulate the pore-scale behavior of single and multiple droplets forming on simplified and realistic porous materials. The comparisons to available experimental and numerical data in the literature show very good agreement for the drop growth behavior. Since in the literature, a wide variation of the presented separation lines has been found, further investigations are recommended to determine the detachment criterion. The criterion developed for this work is a (semi-)analytic approach which can be easily adapted if further information on the force balance composition is available. \\
	The presented model is capable to model multi-drop interactions and confirms the bulging menisci model. Also the flow behavior inside the GDL of receding menisci during the drop occurrence has been successfully captured. Modelling GDL structures is difficult due to the high heterogeneity and complex network structure. The pore-network model is an efficien model that takes the pore-local properties of the GDL into account. However, to evaluate a large sample of the GDL structur (e.g. a representative volume of a GDL structure), still a large amount of memory and computation power is requested and the  numerical stability of the developed model has to be improved. Therefore, the calculation needs to be made even more efficient. Nevertheless, a small sample of a fibrous GDL structure could be represented and the pore-local flow behavior has been analysed. Even though the detailed local behavior has to be handled with care due to pressure fluctuations and numerical inaccuracies in such a complex network, the developed model is capable to represent the general behavior during droplet formation, growth and detachment and presents reasonable and validated results.\\
	The developed model allows to locally analyse the water transport through the GDL in a fuel cell to analyse the flooding behavior of a specific structure and the estimated liquid water outtake by droplets. The presented model can be extended to capture component transport and phase change in the pore-network and at the drop surface. In future work, the channel flow will be directly coupled to the pore-network to account for mass coupling between the two domains including drop occurrence.\\
	This work is part of a project to describe the water transport phenomena at the interface between GDL and gas distributor in the cathode of a PEM fuel cell. Here, not only drop occurrence but also wettability changes at the interface between the two domains play an important role and have a major influence on the water outtake from the GDL structure.
	\section{Acknowledgements}
	We thank Larissa Brencher for providing the ANSYS Fluent model setup that was created during a working student activity at Robert Bosch GmbH. Additionally, we thank the Robert Bosch GmbH for the financial and professional support and the Deutsche Forschungsgemeinschaft (DFG, German Research Foundation) for supporting this work by funding SFB 1313, Project Number 327154368.
	\newpage
	\bibliographystyle{ieeetr}
	\bibliography{new-paper-droplets}

\begin{thebibliography}{10}

\bibitem{ackermann2021multi2}
S.~Ackermann, {\em A multi-scale approach for drop/porous-medium interaction}.
\newblock Stuttgart: Eigenverlag des Instituts f{\"u}r Wasser- und
  Umweltsystemmodellierung, 2021.

\bibitem{barbir2012pem}
F.~Barbir, {\em {PEM} fuel cells: theory and practice}.
\newblock Academic press, 2012.

\bibitem{andersson2019modeling}
M.~Andersson, V.~Vuk{\v{c}}evi{\'c}, S.~Zhang, Y.~Qi, H.~Jasak, S.~Beale, and
  W.~Lehnert, ``Modeling of droplet detachment using dynamic contact angles in
  polymer electrolyte fuel cell gas channels,'' {\em {I}nternational {J}ournal
  of {H}ydrogen {E}nergy}, vol.~44, no.~21, pp.~11088--11096, 2019.

\bibitem{kumbur2009fuel}
E.~Kumbur and M.~Mench, ``Fuel cells - proton-exchange membrane fuel cells
  water management,'' in {\em Encyclopedia of Electrochemical Power Sources}
  (J.~Garche, ed.), pp.~828--847, Amsterdam: Elsevier, 2009.

\bibitem{wang2017effect}
C.~Wang, Q.~Zhang, J.~Lu, S.~Shen, X.~Yan, F.~Zhu, X.~Cheng, and J.~Zhang,
  ``Effect of height/width-tapered flow fields on the cell performance of
  polymer electrolyte membrane fuel cells,'' {\em International Journal of
  Hydrogen Energy}, vol.~42, no.~36, pp.~23107--23117, 2017.

\bibitem{hussaini2009visualization}
I.~S. Hussaini and C.-Y. Wang, ``Visualization and quantification of cathode
  channel flooding in {PEM} fuel cells,'' {\em Journal of Power Sources},
  vol.~187, no.~2, pp.~444--451, 2009.

\bibitem{qin2019dynamic}
C.-Z. Qin, B.~Guo, M.~Celia, and R.~Wu, ``Dynamic pore-network modeling of
  air-water flow through thin porous layers,'' {\em Chemical Engineering
  Science}, vol.~202, pp.~194--207, 2019.

\bibitem{niu2018two}
Z.~Niu, Z.~Bao, J.~Wu, Y.~Wang, and K.~Jiao, ``Two-phase flow in the
  mixed-wettability gas diffusion layer of proton exchange membrane fuel
  cells,'' {\em Applied Energy}, vol.~232, pp.~443--450, 2018.

\bibitem{niblett2020two}
D.~Niblett, A.~Mularczyk, V.~Niasar, J.~Eller, and S.~Holmes, ``Two-phase flow
  dynamics in a gas diffusion layer-gas channel-microporous layer system,''
  {\em Journal of Power Sources}, vol.~471, p.~228427, 2020.

\bibitem{sakaida2017large}
S.~Sakaida, Y.~Tabe, and T.~Chikahisa, ``Large scale simulation of liquid water
  transport in a gas diffusion layer of polymer electrolyte membrane fuel cells
  using the lattice boltzmann method,'' {\em Journal of Power Sources},
  vol.~361, pp.~133--143, 2017.

\bibitem{zhang2018three}
D.~Zhang, Q.~Cai, and S.~Gu, ``Three-dimensional lattice-boltzmann model for
  liquid water transport and oxygen diffusion in cathode of polymer electrolyte
  membrane fuel cell with electrochemical reaction,'' {\em Electrochimica
  Acta}, vol.~262, pp.~282--296, 2018.

\bibitem{straubhaar2015water}
B.~Straubhaar, J.~Pauchet, and M.~Prat, ``Water transport in gas diffusion
  layer of a polymer electrolyte fuel cell in the presence of a temperature
  gradient,'' {\em International Journal of Hydrogen Energy}, vol.~40, no.~35,
  pp.~11668--11675, 2015.

\bibitem{aghighi2017pore}
M.~Aghighi and J.~Gostick, ``Pore network modeling of phase change in {PEM}
  fuel cell fibrous cathode,'' {\em Journal of Applied Electrochemistry},
  vol.~47, no.~12, pp.~1323--1338, 2017.

\bibitem{mularczyk2020droplet}
A.~Mularczyk, Q.~Lin, M.~J. Blunt, A.~Lamibrac, F.~Marone, T.~J. Schmidt, F.~N.
  B{\"u}chi, and J.~Eller, ``Droplet and percolation network interactions in a
  fuel cell gas diffusion layer,'' {\em Journal of The Electrochemical
  Society}, vol.~167, no.~8, p.~084506, 2020.

\bibitem{santini2013x}
M.~Santini, M.~Guilizzoni, and S.~Fest-Santini, ``X-ray computed
  microtomography for drop shape analysis and contact angle measurement,'' {\em
  Journal of colloid and interface science}, vol.~409, pp.~204--210, 2013.

\bibitem{ackermann2021multi}
S.~Ackermann, C.~Bringedal, and R.~Helmig, ``Multi-scale three-domain approach
  for coupling free flow and flow in porous media including droplet-related
  interface processes,'' {\em Journal of Computational Physics}, vol.~429,
  p.~109993, 2021.

\bibitem{weishaupt2019efficient}
K.~Weishaupt, V.~Joekar-Niasar, and R.~Helmig, ``An efficient coupling of free
  flow and porous media flow using the pore-network modeling approach,'' {\em
  Journal of Computational Physics: X}, vol.~1, p.~100011, 2019.

\bibitem{Geodict}
{Math2Market GmbH}, ``Geodict 2018 graingeo.''

\bibitem{gostick2019porespy}
J.~T. Gostick, Z.~A. Khan, T.~G. Tranter, M.~D. Kok, M.~Agnaou, M.~Sadeghi, and
  R.~Jervis, ``Porespy: A python toolkit for quantitative analysis of porous
  media images,'' {\em Journal of Open Source Software}, vol.~4, no.~37,
  p.~1296, 2019.

\bibitem{blunt2017multiphase}
M.~J. Blunt, {\em Multiphase flow in permeable media: A pore-scale
  perspective}.
\newblock Cambridge University Press, 2017.

\bibitem{weishaupt2021dynamic}
K.~Weishaupt and R.~Helmig, ``A dynamic and fully implicit non-isothermal,
  two-phase, two-component pore-network model coupled to single-phase free flow
  for the pore-scale description of evaporation processes,'' {\em Water
  Resources Research}, vol.~57, no.~4, p.~e2020WR028772, 2021.

\bibitem{joekar2012analysis}
V.~Joekar-Niasar and S.~Hassanizadeh, ``Analysis of fundamentals of two-phase
  flow in porous media using dynamic pore-network models: A review,'' {\em
  Critical reviews in environmental science and technology}, vol.~42, no.~18,
  pp.~1895--1976, 2012.

\bibitem{qin2015water}
C.~Qin, ``Water transport in the gas diffusion layer of a polymer electrolyte
  fuel cell: dynamic pore-network modeling,'' {\em Journal of the
  Electrochemical Society}, vol.~162, no.~9, p.~F1036, 2015.

\bibitem{oren1998extending}
P.-E. Oren, S.~Bakke, and O.~J. Arntzen, ``Extending predictive capabilities to
  network models,'' {\em SPE journal}, vol.~3, no.~04, pp.~324--336, 1998.

\bibitem{baber2012numerical}
K.~Baber, K.~Mosthaf, B.~Flemisch, R.~Helmig, S.~M{\"u}thing, and B.~Wohlmuth,
  ``Numerical scheme for coupling two-phase compositional porous-media flow and
  one-phase compositional free flow,'' {\em The IMA Journal of Applied
  Mathematics}, vol.~77, no.~6, pp.~887--909, 2012.

\bibitem{chen2020fully}
S.~Chen, C.~Qin, and B.~Guo, ``Fully implicit dynamic pore-network modeling of
  two-phase flow and phase change in porous media,'' {\em Water Resources
  Research}, vol.~56, no.~11, p.~e2020WR028510, 2020.

\bibitem{Kochetal2020Dumux}
T.~Koch, D.~Gl{\"a}ser, K.~Weishaupt, S.~Ackermann, M.~Beck, B.~Becker,
  S.~Burbulla, H.~Class, E.~Coltman, S.~Emmert, T.~Fetzer, C.~Gr{\"u}ninger,
  K.~Heck, J.~Hommel, T.~Kurz, M.~Lipp, F.~Mohammadi, S.~Scherrer,
  M.~Schneider, G.~Seitz, L.~Stadler, M.~Utz, F.~Weinhardt, and B.~Flemisch,
  ``{DuMu\textsuperscript{x} 3 - an open-source simulator for solving flow and
  transport problems in porous media with a focus on model coupling},'' {\em
  Computers \& Mathematics with Applications}, 2020.

\bibitem{kumbur2006liquid}
E.~Kumbur, K.~Sharp, and M.~Mench, ``Liquid droplet behavior and instability in
  a polymer electrolyte fuel cell flow channel,'' {\em Journal of Power
  Sources}, vol.~161, no.~1, pp.~333--345, 2006.

\bibitem{chen2005simplified}
K.~S. Chen, M.~A. Hickner, and D.~R. Noble, ``Simplified models for predicting
  the onset of liquid water droplet instability at the gas diffusion layer/gas
  flow channel interface,'' {\em International Journal of Energy Research},
  vol.~29, no.~12, pp.~1113--1132, 2005.

\bibitem{cho2012droplet}
S.~C. Cho, Y.~Wang, and K.~S. Chen, ``Droplet dynamics in a polymer electrolyte
  fuel cell gas flow channel: Forces, deformation, and detachment. i:
  Theoretical and numerical analyses,'' {\em Journal of power sources},
  vol.~206, pp.~119--128, 2012.

\bibitem{quesnel2015dynamic}
C.~Quesnel, R.~Cao, J.~Lehr, A.-M. Kietzig, A.~Z. Weber, and J.~T. Gostick,
  ``Dynamic percolation and droplet growth behavior in porous electrodes of
  polymer electrolyte fuel cells,'' {\em The Journal of Physical Chemistry C},
  vol.~119, no.~40, pp.~22934--22944, 2015.

\bibitem{ransohoff1988laminar}
T.~Ransohoff and C.~Radke, ``Laminar flow of a wetting liquid along the corners
  of a predominantly gas-occupied noncircular pore,'' {\em Journal of colloid
  and interface science}, vol.~121, no.~2, pp.~392--401, 1988.

\bibitem{Zhou1997}
D.~Zhou, M.~Blunt, and F.~Orr, ``Hydrocarbon drainage along corners of
  noncircular capillaries,'' {\em Journal of Colloid and Interface Science},
  vol.~187, no.~1, pp.~11 -- 21, 1997.

\bibitem{weishaupt2020model}
K.~Weishaupt, {\em Model concepts for coupling free flow with porous medium
  flow at the pore-network scale: from single-phase flow to compositional
  non-isothermal two-phase flow}.
\newblock Stuttgart: Eigenverlag des Instituts f{\"u}r Wasser-und
  Umweltsystemmodellierung, 2020.

\bibitem{hartnett1989heat}
J.~P. Hartnett and M.~Kostic, ``Heat transfer to newtonian and non-newtonian
  fluids in rectangular ducts,'' in {\em Advances in heat transfer}, vol.~19,
  pp.~247--356, Elsevier, 1989.

\end{thebibliography}
	\newpage
	\section*{Appendix}
	\subsection*{Formulation of the throat conductance factors}
	An expression for the wetting layer conductance is given by Ransohoff and Radke \cite{ransohoff1988laminar},
	\begin{linenomath*}
	\begin{align}
		k_{ij}^w= \frac{r^2_{AM}}{\mu_w l_{ij}} \sum_{n=1}^{n_{corner}} \frac{A_{w,n}}{\beta_n}\,,
	\end{align}
	\end{linenomath*}
	with the dynamic viscosity of the wetting phase $\mu_w$, the throat length $l_{ij}$, the cross-sectional area of the wetting layer in the corner $n$ $A_{w,n}$ and the dimensionless resistance factor $\beta_n$. Where $r_{AM}= \sigma/p_c$ is the radius of curvature of the arc menisci (AM) an $n_{corner}$ the number of corners of the throat shape. The resistance factor can be determined numerically by solving the stationary Stokes equation for a single wetting layer. It depends on the corner geometry and the phase configuration (corner angle and capillary pressure). However, we use am analytical approximation proposed by Zhou et al.\cite{Zhou1997}
	\begin{linenomath*}
	\begin{align*}
		\beta_n = \frac{ 12 \sin^2(\alpha_n)\left(1-B \right)^2}{\left(1-\sin(\alpha_n)\right)^2 B^2} \cdot 
		\frac{\left(\phi_1 - B\phi_2 \right) \left(\phi_3 + f_2 B \phi_2 - \left(1-f_1 B\right)r_{ij}\right)^2}{\left(\phi_1 - B \phi_2 -\left(1-B \right) r_{ij}^2 \right)^3}
	\end{align*}
	\end{linenomath*}
	with
	\begin{linenomath*}
	\begin{align*}
		B &= \left(\frac{\pi}{2}-\alpha\right)\tan(\alpha)\\
		\phi_1 &= \cos^2(\alpha + \theta)+ \cos(\alpha + \theta)\sin(\alpha + \theta)\tan(\alpha) \\
		\phi_2 &= 1-\frac{\theta}{\pi/2 -\alpha} \\
		\phi_3 &= \frac{\cos(\alpha + \theta)}{\cos(\alpha)}
	\end{align*}
	\end{linenomath*}
	where $\alpha_n$ is the half corner angle of corner $n$ and $\theta$ is the contact angle in corner $n$. The factors $f_1$ and $f_2$ correspond to the interface conditions between the phases. Dependent on the fluid viscosity ratio, a no-slip ($f_2 = 1$), a no-stress ($f_2=0$) or an intermediate condition ($0<f_2<1$) may be assigned. For a water-air-solid setup, a no-slip condition can be applied \cite{blunt2017multiphase}. For the interface between the solid and the wetting phase, a no-slip condition is assumed and thus $f_1=1$ \cite{weishaupt2020model}. \\
	For the flow of the non-wetting phase, we use the following expression proposed by Blunt \cite{blunt2017multiphase}
	\begin{linenomath*}
	\begin{align}
		k_{ij}^n = \frac{c A_{n,ij}^2 G}{\mu_n l_{ij}}\,,
	\end{align}
	\end{linenomath*}
	with the non-wetting phase dynamic viscosity $\mu_n$, the actual cross-sectional area of the non-wetting phase $A_{n,ij}$ and the shape factor $G$ and proportionality constant $c$ dependent on the cross-sectional area shape. For circles it takes the value $0.5$, for squares $c=0.5623$ and for equilateral triangles $c = 0.6$.
	\subsection*{Derivation of the capillary pressure saturation relation for droplets}
\begin{figure}[h!]
	\centering
	\setlength\figureheight{5cm}
	\setlength\figurewidth{10cm}
	\input{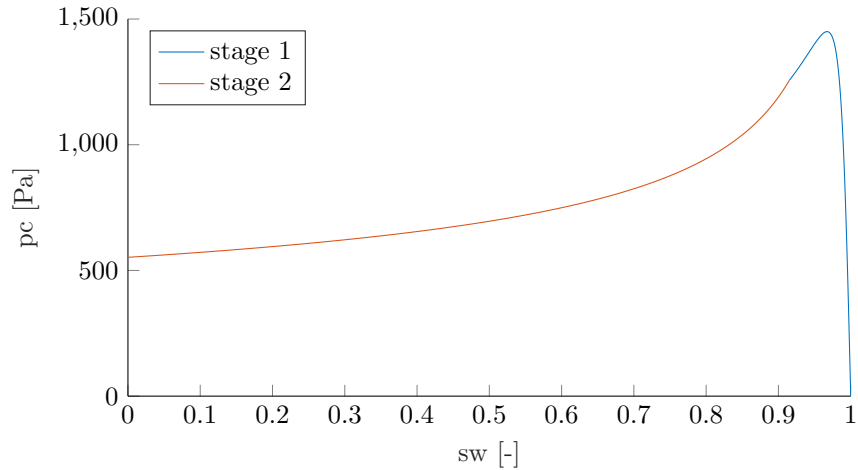}
	\caption{Composed local capillary pressure saturation relation for droplet pores.}
	\label{fig:pc-sw-drop}
\end{figure}

	The pressure in the droplet is calculated using a local capillary pressure saturation relation analog to the pore-network concept. The shape of the droplet and the resulting curvature is calculated using two different formulations dependent on the size/state of the drop.\\
It is assumed that the drop emerges from a throat. Before a droplet with the according contact angle is formed on the hydrophobic surface, a sphere cap is developed with increasing contact angle until the static contact angle of the material is reached. During this state, the contact area of the drop is constant and equals the cross-sectional area of the throat. This first stage is visualized in Fig. \ref{fig:dropStages} (left). Once the final contact angle between drop and surface is reached, it stays constant during the further growing. Instead, the droplet growth causes a change in the contact area of the droplet with the hydrophobic surface. The second stage of droplet growth is visualized in Fig.\ref{fig:dropStages} (right).
\begin{figure}[h!]
	\def\svgwidth{\linewidth}
	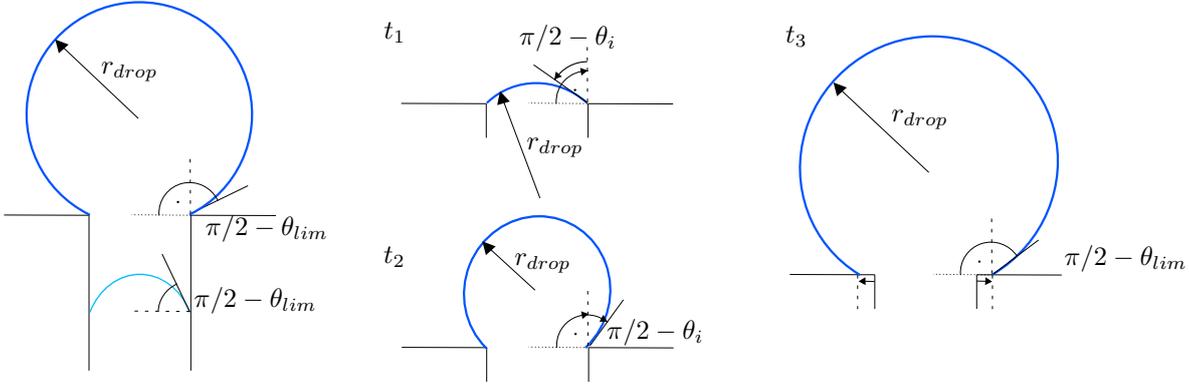
	\caption{Different stages of a growing droplet ($t_1 < t_2 < t_3$).}
	\label{fig:DropletGrowth-Angles}
\end{figure}
The capillary pressure saturation formulation for stage 1 is derived from the volume of a sphere cap and its corresponding radius/curvature
\begin{linenomath*}
\begin{align}
	V_{drop,1} = \frac{\pi}{3} r_{curv,i}^3\left(2+\cos(\pi/2-\theta_i)\right)\left(1-\cos(\pi/2-\theta_i)\right)^2\,,
\end{align}
\end{linenomath*}
with the curvature radius $r_{curv}$, and the changing contact angle $\theta_i \in \left[\pi/2, \pi/2-\theta_{lim}\right]$ with $\theta_{lim} = pi-\theta_a$. The curvature radius is calculated based on the constant contact area $A_{CA} = A_{throat}$, 
\begin{linenomath*}
\begin{align}
	r_{curv,i} = \frac{r_{throat}}{\cos(\theta_i)}\,.
\end{align}
\end{linenomath*}
The resulting capillary pressure formulation for stage 1 is
\begin{linenomath*}
\begin{align}
	p_{c,1}=\frac{2\sigma\cos(\theta_i)}{r_{throat}}\,.
\end{align}
\end{linenomath*}
If the drop pore is not invaded yet, the capillary pressure is zero.\\
The corresponding saturation in the drop pore is given by the drop volume divided by the pore volume
\begin{linenomath*}
\begin{align}
	s_{w,1} = 1 - s_n = 1- \frac{V_{drop,1}}{V_{pore}}\,.
\end{align}
\end{linenomath*}
In stage 2, the contact angle is constant and equals the advancing contact angle of the fluid on the hydrophobic surface ($\theta_i \equiv \theta_{lim}$). Now, the contact area is changing.
\begin{linenomath*}
\begin{align}
	V_{drop,2} = \frac {\pi}{3}r_{drop,i}^3\left(2+\cos(\pi/2-\theta_{lim})\right)\left(1-\cos(\pi/2-\theta_{lim})\right)^2\,,
\end{align}
\end{linenomath*}
with the drop radius $r_{drop,i} \in \left[r_{throat}/\cos(\theta_{lim}),1.32r_{pore}\right]$ (1.32 due to $s_w \rightarrow 0$ to complete the function (does not occur))
The resulting capillary pressure formulation for stage 2 is
\begin{linenomath*}
\begin{align}
p_{c,2}=\frac{2\sigma\cos(\theta_i)}{r_{drop}}\,.
\end{align}
\end{linenomath*}
Again, the  saturation in the drop pore is given by the drop volume devided by the pore volume
\begin{linenomath*}
\begin{align}
s_{w,2} = 1 - s_n = 1- \frac{V_{drop,2}}{V_{pore}}\,.
\end{align}
\end{linenomath*}
The complete, local capillary pressure saturation for drop pores is visualized in Fig.\ref{fig:pc-sw-drop}.
	\subsection*{Derivation of the drag force based on control volumes}
	\subsubsection*{Simplified droplet in a channel}
\begin{figure}[h!]
	\centering
	\def\svgwidth{0.8\linewidth}
	\input{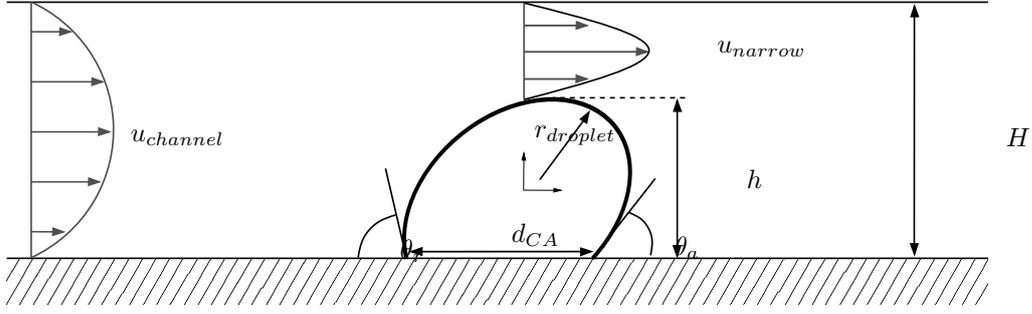}
	\caption{Simplified representation of a droplet on the GDL surface}
	\label{fig:SimplifiedDroplet}
\end{figure}

In Fig.\ref{fig:SimplifiedDroplet}, the considered setup of a droplet in a channel is shown.
\subsubsection*{Force balance at the droplet control volume}
\begin{figure}[h!]
	\centering
	\def\svgwidth{0.8\linewidth}
	\input{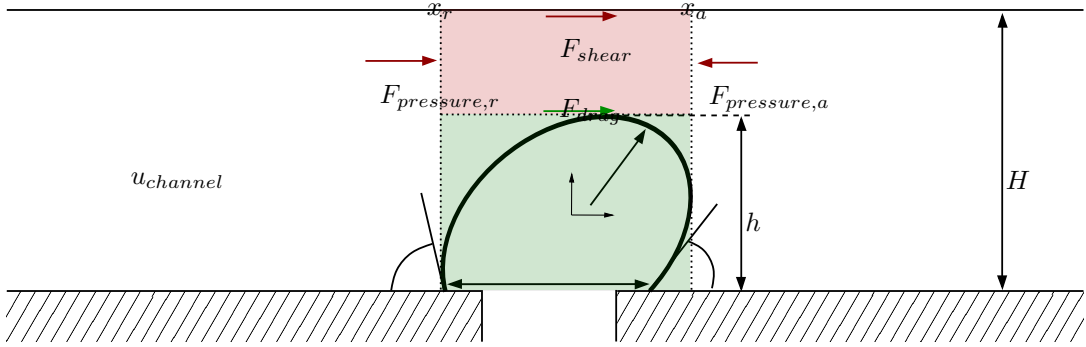}
	\caption{Force balance on the separated control volume to estimate the drag force in the droplet control volume}
	\label{fig:SimplifiedDropletForces}
\end{figure}

\begin{figure}
    \centering
    \includegraphics[width = 0.8 \textwidth]{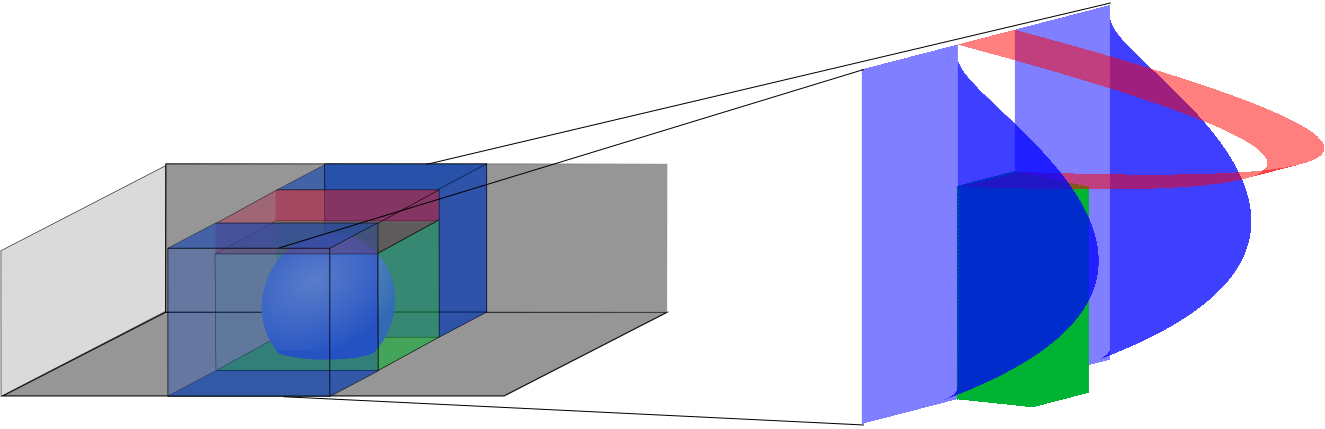}
    \caption{Three-dimensional control volume approach: Extension of the concept presented by Kumbur et al.\cite{kumbur2006liquid} by the blue CVs to calculate the side shear forces which result from zero velocity assumption at the drop surface/CV surface.}
    \label{fig:Droplet3DCVs}
\end{figure}

Several different forces are acting on the droplet in the channel flow. Since we are mainly interested in the detachment behavior, only the force balance in the streamwise direction (x-direction) is considered, here (see Fig.\ref{fig:SimplifiedDropletForces}). 
We follow an approach similar to the one presented by Kumbur et al. \cite{kumbur2006liquid} but place additional control volumes (blue) next to the drop CV (green), as shown in Fig.\ref{fig:Droplet3DCVs}. On the left side, the position and size of the CVs is shown. On the right side, the approximated velocity profiles around the droplet are visualized.
\\ The force balance is given by:
\begin{linenomath*}
\begin{align*}
	F_{drag} + F_{shear} + F_{pressure} = 0 \,.
\end{align*}
\end{linenomath*}
\paragraph*{Calculation of the pressure force}
The pressure force acting on the droplet control volume (green CV) is given following \cite{kumbur2006liquid} by
\begin{linenomath*}
\begin{align*}
	F_{pressure} = \frac{24 \mu_\text{gas} \left(H/2\right)^2 u_\text{channel} h_\text{drop}^2}{\left(H/2 -h_\text{drop}/2\right)^3\left(1-\cos\theta_a\right)^2} \qquad \text{where} \qquad r_\text{drop} = \frac{h_\text{drop}}{1-\cos \theta_a}\,,
\end{align*}
\end{linenomath*}
with the channel height $H$, the droplet height $h_\text{drop}$, the droplet radius $r_\text{drop}$, the mean velocity in the channel $u_\text{channel}$, the dynamic viscosity of air $\mu_\text{gas}$ and the advancing contact angle $\theta_a$.
\paragraph*{Calculation of the shear force}
A cubic control volume is considered for the droplet. A shear force is acting on three sides of the droplets: On the top and the two sides parallel to the flow direction. The droplet and its surrouding conditions are assumed to be symmetric in the channel. Therefore, the shear force is calculated in two parts
\begin{linenomath*}
\begin{align*}
	F_{shear} = F_{shear}^{top} + 2 F_{shear}^{side}\,.
\end{align*}
\end{linenomath*}
The shear force at the top of the droplet CV is calculated following Kumbur et al. \cite{kumbur2006liquid} by considerng the shear force in the streamwise (x) direction at the upper wall of the control volume
\begin{linenomath*}
\begin{align*}
	F_{shear}^{top} = \left. \tau \right|_{y'=b/2} A = \frac{3 \mu_\text{gas} H/2 u_\text{channel}}{\left(H/2 -h_\text{drop}/2\right)^2} \left(2r_\text{drop}\right)^2\,,
\end{align*}
\end{linenomath*}
with the shear stress $\tau$ and the projected area of the droplet $A$. Here, a quadratic approximation is applied for the area.\\
To get the shear force acting on the sides of the droplet CV, the velocity profile of the flow next to the droplet needs to be taken into account. We assume a fully developed, laminar flow of a Newtonian fluid, such that a parabolic flow profile is established. Additionally, we assume a no-slip condition at the droplet surface (CV surface) and the channel walls.\\
With the z-direction normal to the droplet CV face, perpendicular to the stream, the following expression for the velocity profile is given based on the flow between two parallel plates
\begin{linenomath*}
\begin{align*}
	u(z) = \frac{3}{2} U \left[1-\left(\frac{2z}{H}\right)^2\right]\,.
\end{align*}
\end{linenomath*}
From this velocity profile, we know:
\begin{enumerate}
	\item The maximum shear stress results in the middle of the channel.
	\item The shear stress is zero in the lower edge (contact of droplet and channel wall).
	\item A linear relation can be applied for the shear stress $\tau$.
\end{enumerate}
The shear stress at the sides of the droplet CV is then given by
\begin{linenomath*}
\begin{align*}
	F_{shear}^{side} = \left. \tau_{yz} \right|_{z=r} A = -\mu_\text{gas} \int_{y=-H/2}^{y=-H/2+h_\text{drop}} \frac{\partial u}{\partial z}(y) dy \cdot 2r_\text{drop} = \frac{12 \mu_\text{gas} r_\text{drop} h_\text{drop} u_\text{channel}}{H}\left(2-\frac{h_\text{drop}}{H} \right)\,,
\end{align*}
\end{linenomath*}
with the velocity gradient perpendicular to the CV surface and the streamflow dependent on the position in the channel $\frac{\partial u}{\partial z}(y)$. Since the velocity gradient changes dependent on the y-direction in the channel, we integrate over the height of the droplet $h_\text{drop}$. For the size of the droplet in streamwise (x) direction $2r_\text{drop}$ is assumed and the surface is approximated by a rectangle $A=h_\text{drop}\cdot 2r_\text{drop}$.\\
To take the reduction of the cross-sectional area due to the droplet into account, which results in an increased velocity at the sides of the droplets, a correction factor is introduced
\begin{linenomath*}
\begin{align}
	f_\text{channel} = \left(3/2\frac{A_\text{channel}}{A_\text{channel sides}}\right)^2 =  \left(3/2\frac{A_\text{channel}}{A_\text{channel}-2r_\text{drop}H}\right)^2\,,
\end{align}
\end{linenomath*}
	dependent on the cross-sectional area of the whole channel $A_\text{channel}$ and the part of the cross-section which is not blocked by the droplet $A_\text{channel sides}$.\\
	The multiplier $3/2$ results from the parabolic flow profile, where the maximum velocity in the channel is $3/2$ times the average velocity. The exponent 2 is used since the channel is limited in height and width.
\paragraph*{Resulting drag force}
The resulting drag force acting on the droplet CV is calculated from the force balance
\begin{linenomath*}
\begin{align*}
	F_{drag} &= -F_{pressure} -F_{shear} = -F_{pressure} -F_{shear}^{top} - 2F_{shear}^{side} \\
	&= - \frac{24 \mu_\text{gas} \left(H/2 \right)^2 u_\text{channel} h_\text{drop}^2}{\left(H/2-h_\text{drop}/2\right)^3\left(1-\cos\theta_a\right)^2} - \frac{3\mu_\text{gas} H/2 u_\text{channel}}{\left(H/2 -h_\text{drop}/2 \right)^2}\left(2r_\text{drop}\right)^2 - 2\cdot f_\text{channel} \frac{12 \mu_\text{gas} r_\text{drop} h_\text{drop} u_\text{channel}}{H}\left(2-\frac{h_\text{drop}}{H}\right) \,.
\end{align*}
\end{linenomath*}
This gives a linear relation for the drag force on the average velocity in the channel $U$.
	\subsection*{Model setup of the ANSYS Fluent simulation}
    In the following an overview on the ANSYS Fluent model setup is given.
	\begin{figure}[h!]
	\centering
	\includegraphics[width=0.9\textwidth]{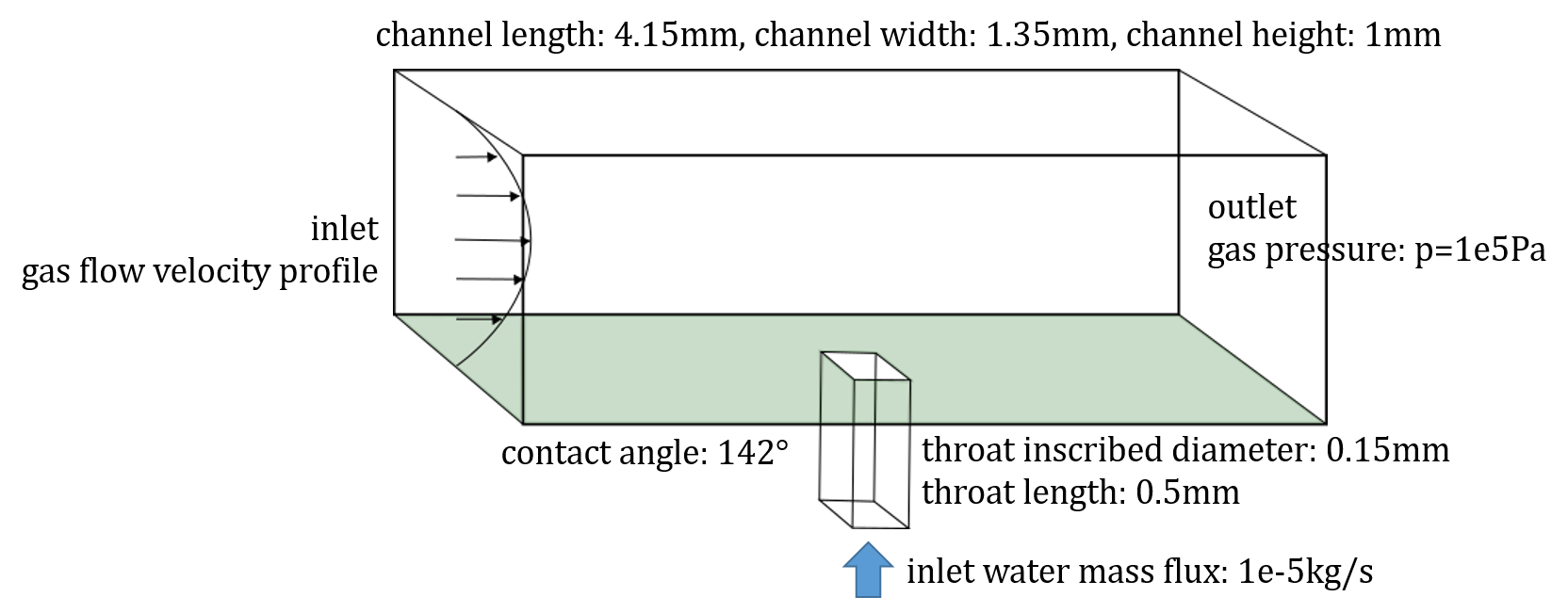}
	\caption{Setup of the CFD simulation performed using Ansys Fluent R2019. Gas channel with parabolic gas flow velocity inlet and pressure outlet. In the center, water is injected through a throat with a mass flux boundary conditions. The drops form inside the channel and are deformed and detached by the surrounding gas flow.}
	\label{fig:AnsysSetup}
	\end{figure}
	A channel of length $L=4.15mm$, width $W=1.35mm$ and height $H=1mm$ has been analysed with a two-phase VoF model without phase change (immiscible phases) as implemented in ANSYS Fluent version R2019. A filling throat with square cross-sectional area is connected to the channel as shown in Fig.\ref{fig:AnsysSetup}. The square cross-sectional shape of the filling throat is chosen to simplify the meshing of the domain. Since a mass flux boundary condition is applied to the complete inlet face of the throat (volume fraction of water equals 1), no wetting phase flow (corner flow) can form in this throat and the square shape has only a minor influence on the drop formation. The equilibrium contact angle of water with the solid surface is $142^\circ$ at the bottom of the channel (green in Fig.\ref{fig:AnsysSetup}) and the filling throat which represents GDL properties. Hydrophilic surfaces are applied to the other channel walls such that the liquid water is attached to the wall in case of contact. This has no influence on the drop formation, growth and detachment in the considered cases. No contact angle hysteresis is implemented. The droplet deformation results from local forces at the drop surfaces. At the channel inlet, air is injected into the channel with a fully developed laminar flow with a parabolic velocity profile \cite{hartnett1989heat}:
	\begin{linenomath*}
    \begin{align*}
    v(y,z) = \max(v_{gas})^{inlet} \cdot \left(1-\left(\frac{\left|y-a\right|}{a}\right)^2\right)\cdot\left(1-\left(\frac{\left|z\right|}{b}\right)^{2.3}\right)
    \end{align*}
    \end{linenomath*}
    where, $a$ and $b$ are half of the width and height of the channel, respectively. Atmospheric conditions (pressure boundary condition) are applied at the channel outlet and water back-flow of the liquid phase is permitted in the ANSYS Fluent model. \\
    Water is injected into the throat at a constant mass flow rate of $\Dot{m}=10^{-5} kg/s$. Droplets form and grow in the channel filled by the throat. Finally, the droplets detach due to the forces resulting from the gas flow in the channel. 
\end{document}